\documentclass[aps,twocolumn]{revtex4}%prx,color,psfig,epsf
\usepackage[utf8]{inputenc}
\usepackage{graphicx}
\usepackage{amsmath}
\usepackage{mathtools}
\usepackage{amssymb}
\usepackage{color}
\usepackage{amsfonts}
\usepackage{epsf}
\usepackage{epstopdf}
\usepackage{bm}

\allowdisplaybreaks

\usepackage{hyperref}
\def\sectionautorefname~#1\null{Section~(#1)\null}
\def\equationautorefname~#1\null{Eq.~(#1)\null}
\def\figureautorefname~#1\null{Fig.~#1\null}
\def\tableautorefname~#1\null{Table~#1\null}

\newcommand{\ti}[1]{\tilde{#1}}
\newcommand{\sdim}[1]{\left[#1\right]}
\newcommand{\avg}[1]{\left\langle#1\right\rangle}

\begin{document}

\title{A renormalization group study of the dynamics of active membranes: universality classes and scaling laws}
\author{Francesco Cagnetta,  Viktor \v{S}kult\'{e}ty, Martin R. Evans, and  Davide Marenduzzo}
\affiliation{SUPA, School of Physics and Astronomy, The University of Edinburgh, Edinburgh, EH9 3FD, Scotland, United Kingdom}

\begin{abstract}
Motivated by experimental observations of patterning at the leading edge of motile eukaryotic cells, we introduce a general model for the dynamics of nearly-flat fluid membranes driven from within by an ensemble of activators. We include, in particular, a kinematic coupling between activator density and membrane slope which generically arises whenever the membrane has a non-vanishing normal speed. We unveil the phase diagram of the model by means of a perturbative field-theoretical renormalization group analysis. Due to the aforementioned kinematic coupling the natural early-time dynamical scaling is acoustic, that is the dynamical critical exponent is $1$. However, as soon as the the normal velocity of the membrane is tuned to zero, the system crosses over to diffusive dynamic scaling in mean field. Distinct critical points can be reached depending on how the limit of vanishing velocity is realised: in each of them corrections to scaling due to nonlinear coupling terms must be taken into account. The detailed analysis of these critical points reveals novel scaling regimes wich can be accessed with perturbative methods, together with signs of strong coupling behavior, which establishes a promising ground for further non-perturbative calculations. Our results unify several previous studies on the dynamics of active membrane, while also identifying nontrivial scaling regimes which cannot be captured by passive theories of fluctuating interfaces and are relevant for the physics of living membranes.
\end{abstract}

\maketitle

\section{Introduction}

The study of statics and dynamics of fluid membranes, a classic problem in soft condensed matter, has been rekindled in the context of cell biophysics. This is due to the paramount importance of the plasma membrane of eukaryotic cells in biological processes and the new features it brings to membrane physics.  In contrast with a `passive' fluid membrane, which can be thought of as a $d$-dimensional fluid fluctuating in $d\,{+}\,1$ dimensions, the plasma membrane is an {\em active} membrane,  characterised by the presence of additional active matter, which is embedded and moves within it.  The latter consists mostly of various kinds of membrane proteins that actuate and regulate the many biological processes in which the membrane is involved.

The richness of physical phenomena  generated is remarkable given the relative simplicity of the underlying model system for an active membrane, which can often be described by a set of coupled equations for the membrane height and the density of active proteins~\cite{prost1996}. On the one hand, each of these proteins can be associated with a specific energy-consuming process which, having a distinct effect on the membrane properties, must be accounted for. On the other hand, leitmotifs in the statics and dynamics of living membranes suggest that some crucial properties of these membranes are caused by the proteins' activity {\it per se} rather than by the occurrence of a specific process. For instance, superficially similar transverse waves, akin to those observed on advancing lamellipodia~\cite{allard2013}, are found with models considering either a thermodynamic coupling between protein dynamics and interfacial curvature~\cite{gov2006,veksler2007} or a kinetic coupling with the interfacial height~\cite{ramaswamy2000,maitra2014,cagnetta1, cagnetta2, bisht2019}, and it is important to assess whether the two phenomena are fundamentally different or closely related at a deeper level.

The idea of a deep relation between superficially different problems is formalised in equilibrium statistical physics with the concept of {\it universality class}, and there is a dedicated set of techniques aimed at the identification of such classes: the renormalization group (RG). In fact, RG has proved extremely influential and instrumental in the attainment of a fundamental understanding of the possible types of critical points~\cite{wilson1974,cardyScaling,amitFieldTheory}. For instance, this program of study has shown that phenomena as diverse as the order-disorder transition in an Ising magnet and the liquid-vapor transition of a Lennard-Jones fluid are characterised by the same critical exponents, hence belong to the same universality class. In active systems, an analogous second-order phase transition was first observed in the Vicsek model~\cite{vicsek,ginelli-vicsek}, the RG analysis of which was pioneered in the work by Toner and Tu on their theory of flocking~\cite{toner1998,toner2005}. Since then, universal scaling has been  studied in various active systems, from incompressible polar flocks \cite{Chen2015,Cavagna2018} to models showing motility induced phase separation~\cite{wittkowski2014,Caballero2020,tailleur2008,cates-MIPS}. One of the special features of these nonequilibrium systems is that the dynamical critical exponent, describing coupling between spatial and temporal scales, can attain unusual values\cite{Cavagna2017,Cavagna2019,Cavagna2019a} with respect to standard models of equilibrium statistical physics~\cite{hoenberghalperin-critical}.

Our goal in this work is to apply the framework of field-theoretical renormalization group~\cite{tauberCriticalDynamics,Vasilev04} to a generic model for active membranes, with the aim of classifying its possible behavior into classes of models. In doing so, we also link the active membrane problem to the kinetic roughening literature, which studies the scale-invariant property of passive interfaces, both in and out of equilibrium. In this context, we find that activity results in novel and unexpected scaling behavior, going beyond that observed in passive interface models such as the Edwards-Wilkinson (EW)~\cite{EW} and Kardar-Parisi-Zhang (KPZ)~\cite{KPZ} equations.

Specifically, we analyse the problem of a membrane whose motion is controlled by activators. Our strategy is to derive on theoretical  grounds the most general coupled system of equations describing the evolution of the  membrane height and activator density. We identify a crucial term, which arises geometrically and couples the activator dynamics to the membrane slope. We then apply a renormalization group approach to such equations. We adopt a scheme which is perturbative in nature, which, strictly speaking, gives accurate results only close  to the `naive' upper critical dimension. Importantly, though, for many relevant sets of parameters  this upper critical dimension is $2$, which is the physically relevant dimension of the problem. We find that the scale-invariant properties of the system depend on the exact structure of the activator-membrane interaction, thus generating different possible phases according to which coupling terms are present and relevant. The result is a phase diagram in which each phase is exemplified by a different set of minimal equations. We then derive the universal properties corresponding to each phase in the one-loop approximation.

The paper is structured as follows.
\begin{itemize}
 \item[i)] In~\autoref{sec:aieq} we derive general equations of motion for a membrane driven from within by point-like activators, then discuss the relation of such equations to previous works on active membranes or general fluctuating interfaces and introduce our field-theoretic renormalization group approach;
 \item[ii)] ~\autoref{sec:two} is devoted to the mean-field theory of the model, i.e. the study of the linearised equations of motion. In particular:
 \begin{itemize}
   \item[a)] we perform a stability analysis of the model in \autoref{sec:linstab}, which allows us to identify acoustic and diffusive dynamic scaling regimes,
   \item[b)] we identify four possible regimes of critical behaviour via power counting in~\autoref{sec:powercounting}, thus building the mean-field phase diagram of the model (\autoref{fig:phase-diagram}),
   \item[c)] for each regime (Sections~\ref{ssec:acoustic scaling} to~\ref{ssec:passivesliders}) we compute mean-field critical exponents and determine the `naive' upper critical dimension, above which mean-field precitions should hold exactly;
 \end{itemize}
 \item[iii)] In~\autoref{sec:renormalization-diffusive} we study in detail the first of the diffusive regimes---the `active KPZ' model---whose equations consist of a KPZ equation~\cite{KPZ} for the membrane height coupled quadratically with a diffusing field. By computing one-loop corrections to model's vertex functions around the upper critical dimension $d\,{=}
 \,2$, we extend the roughening transition scenario of the standard KPZ equation~\cite{KPZ} to our active model and find a novel perturbative fixed point below $d\,{=}
 \,2$ (\autoref{fig:phase-diagram-KPZ});
 \item[iv)] In~\autoref{sec:renormalization-diffusive-2} we study the other two diffusive regimes, which we call `curvotactic activators' and `passive sliders' model. For the former, we compute one-loop corrections to vertex functions and find no perturbative fixed point in the coupling parameters space, but signs of strong-coupling behavior. For the latter, we prove non-renormalizability by computing the primitive degree of divergence of the corrections to vertex functions.
\end{itemize}
Finally, \autoref{sec:conclusions} contains a discussion of the results and our conclusions.
%In~\autoref{sec:aieq}, we derive the general equations of motion, discuss their relation to previous works on active membranes and introduce our perturbative field-theoretic renormalization group approach. In \autoref{sec:two} we consider these equations at the mean-field level. We first perform a linear stability analysis in \autoref{sec:linstab}, which allows us to identify acoustic and diffusive scaling regimes. Then, in~\autoref{sec:powercounting}, we perform a power-counting study of the scaling of an active membrane, which allows us to define four main cases, or universality classes. In the remainder of the section we present these universality classes and discuss when our power counting analysis is exact and when it fails, so that a full renormalization group analysis is required. We provide such a study for the active KPZ regime in~\autoref{sec:renormalization-diffusive} and for other diffusive regimes in~\autoref{sec:renormalization-diffusive-2}.

\section{Equations of motion and action for active membranes}
\label{sec:aieq}

In this section we outline the first-principles derivation of hydrodynamic equations for a fluctuating membrane driven by an ensemble of activators. Our derivation reveals the emergence of a coupling between the interface slope and the activator density which results from geometrical considerations~\cite{cai1994,cai1995,ramaswamy2000,maitra2014} and plays a key role in determining the universality classes of the system. We also set up, in~\autoref{ssec:formalism}, the field-theoretical framework which we will be using for our renormalization-group calculations.

\subsection{Derivation of the hydrodynamic equations of motion}

At lengthscales larger than its own thickness, a membrane can be described as a $d$-dimensional manifold in a $d\,{+}\,1$-dimensional space. To avoid ambiguity in vector dimensionality, in this section we we will denote $d\,{+}\,1$-dimensional vectors with capital blackboard-bold symbols ($\mathbb{X}$), whereas vectors in $d$ dimensions are denoted with lowercase bold symbols ($\bm{x}$). The variables describing our system are the membrane configuration and the position of the activators, which reside within the membrane. Activators might span the whole membrane or lie within just one of the two lipid layers: at the scale of our model, where the membrane thickness is negligible, these subtle differences are not resolved. A set of hydrodynamic continuum equations for the membrane and activator densities can be written by combining conservation laws with constitutive equations for the forces acting on the system. This was done, for instance, in~\cite{cai1994,cai1995} for passive fluid membranes, or in~\cite{maitra2014} for membranes coupled to a network of polymerising actin filaments. In this section we retrace the key points of the derivation in~\cite{cai1995}, extending it to the case of an active membrane driven by an ensemble of activators.

We work within the Monge gauge: in the (standard) approximation that the membrane is nearly flat and there are no overhangs, the membrane can be parametrised, for each point $\bm{x}$, with the distance $h(\bm{x})$ from a reference plane, i.e.
\begin{equation}\label{eq:monge-gauge}
\mathbb{X}(\bm{x})=\left(x_1,\dots,x_d,h(\bm{x})\right).
\end{equation}
If we denote with $\{\bm{y}_1,\ldots,\bm{y}_N\}$ the positions of the $N$ activators on the membrane, we can define a coarse-grained activator density as follows,
\begin{equation}\label{eq:act-density}
\rho(\bm{x},t) = \frac{m}{\sqrt{g}}\sum_{n=1}^N \delta(\bm{x}-\bm{y}_n(t)),
\end{equation}
where $m$ denotes the activator mass. Note the appearance of the factor $\sqrt{g}$, with $g$ the determinant of the metric tensor associated with the membrane manifold, which is required to ensure that the activator density is invariant with respect to reparametrisations of the membrane coordinates~\cite{cai1994,cai1995}. In the Monge gauge, $g=1+(\bm{\nabla}h)^2$.

\begin{figure}[t!]
	\centering
	\includegraphics[width=0.5\textwidth]{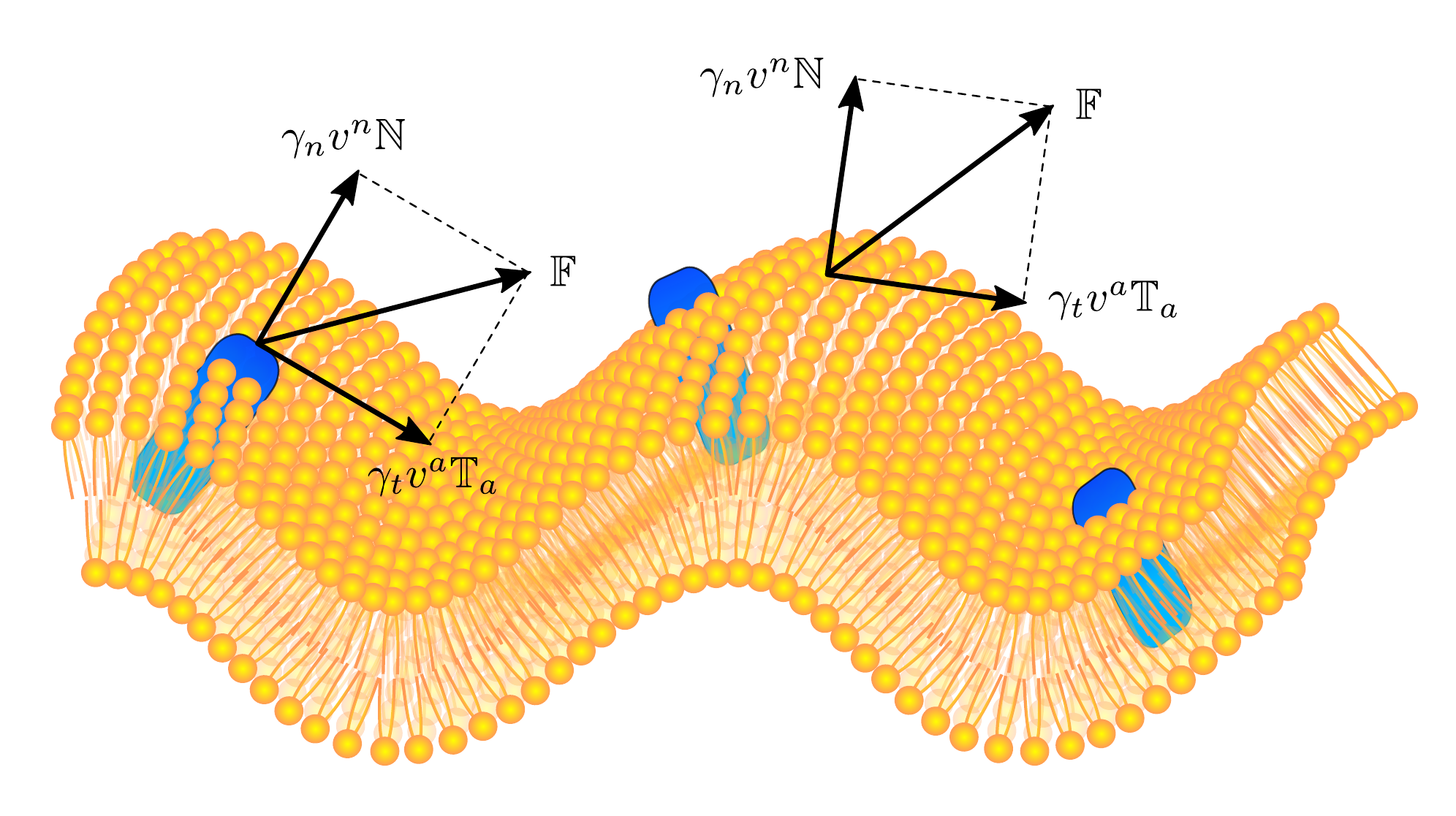}
	\caption{Forces acting on the membrane (yellow) with embedded activators (blue). The vectors $ \gamma_{n} v^{n} \mathbb{N} $ and $ \gamma_{t} v^{a} \mathbb{T}_{a} $ represent the normal and tangential components of the force $ \mathbb{F} $.}
	\label{fig:forces}
\end{figure}

Let us now consider a force acting on the membrane. Any such force can be decomposed into normal and tangential components with respect to the membrane reference plane. In the Monge gauge, the normal direction is given by the following $d+1$-dimensional vector:
\begin{equation}\label{eq:monge-gauge-normal}
 \mathbb{N}(\bm{x}) = \frac{1}{\sqrt{g}}\left(-\partial_1 h,\dots,-\partial_d h ,1\right).
\end{equation}
By assuming overdamped motion for both the interface and the activator proteins within the interface, we can write the following force-velocity relation (or constitutive equation),
\begin{equation}\label{eq:constitutive-eq}
\mathbb{F} = \gamma_n v^n \mathbb{N} + \gamma_t v^a \mathbb{T}_a,
\end{equation}
where $\mathbb{F}$ is the applied force, $\gamma_n$ and $\gamma_t$ are the damping coefficients in the normal and tangential directions, whereas $v^n$ and $v^a$ ($a=1,\ldots,d$) are the components of normal and tangential velocities (see Fig. \ref{fig:forces}). 
Summation over the index $a$ is implied.

Since the activators are constrained to lie within the membrane plane, the only normal motion allowed -- for either membrane patches or activators -- is through dispacement of the membrane. Therefore, in the Monge gauge we have
\begin{equation}\label{eq:membrane-eq}
 v^n = \mathbb{N}\cdot \partial_t \mathbb{X}(\bm{x},t) = \frac{1}{\sqrt{g}}\partial_t h(\bm{x},t). 
\end{equation}
\autoref{eq:membrane-eq} constitutes the equation of motion for the membrane height $h(\bm{x}, t)$.

On the other hand, tangential motion comprises both membrane and activator displacements (see also~\cite{maitra2014}). Therefore, the in-plane activator current, $j^a$, is given by
\begin{equation}\label{eq:activators-eq}
 \rho v^a = j^a + \mathbb{T}^a \cdot \partial_t \mathbb{X}(\bm{x},t) = j^a + \frac{\rho\partial^a h}{g} \partial_t h.
\end{equation}
The first term in the right hand side of \autoref{eq:activators-eq} represents the motion of the activators {\it relative} to the membrane. The second term, instead, describes the motion of the activators {\it due to} the motion of the membrane -- i.e., the membrane motion generates a kinematic force on the activators which is proportional to the membrane slope. The hydrodynamic equation for the activator density $\rho(\bm{x},t)$ has then the form of a covariant conservation law,
\begin{equation}\label{eq:covariant-conservation}
 \sqrt{g}^{-1}\partial_t(\sqrt{g}\rho) + \sqrt{g}^{-1}\partial_a(\sqrt{g} j^a) =0,
\end{equation}
where the flux $j^a$ is given by \autoref{eq:activators-eq}.

In order to complete the model, we must now specify the force $\mathbb{F}$ in the constitutive equation (\ref{eq:constitutive-eq}). We consider the combination of: (i) an active force, directed along the membrane normal and depending on the activator density, (ii) a relaxational force obtained from the functional derivative of a suitable free energy and (iii) stochastic forces whose amplitudes are dictated by the fluctuation-dissipation theorem. Namely, in the Monge gauge,
\begin{equation}\label{eq:constitutive-eq-model}\begin{aligned}
 \mathbb{F} =& \left(f(\rho) -\frac{\delta \mathcal{F}[h,\rho]}{\delta h}  + \sqrt{2k_B T_n\gamma_n}\xi_n(\bm{x},t)\right)\mathbb{N} \\ +& \left(-\rho\partial^a \frac{\delta \mathcal{F}[h,\rho]}{\delta \rho} - \sqrt{2k_BT_t\gamma_t\rho}\xi^a(\bm{x},t)\right) \mathbb{T}_a,
\end{aligned}\end{equation}
where $f(\rho)$ is the modulus of the active force, $\mathcal{F}[h,\rho]$ the free energy, $k_B$ the Boltzmann constant, $T_t$ and $T_n$ the temperatures of membrane and embedding medium, and $\xi_n$ and $\xi_a$ are independent space-time white Gaussian noises with zero mean and unitary variance. Let us stress that active forces acting in the membrane's tangent plane can also be considered and might be relevant in the context of cellular uptake~\cite{chen2020} and shape control~\cite{alimohamadi2020non}.

A comparison with \autoref{eq:constitutive-eq} shows that the two terms in brackets on the right-hand side of \autoref{eq:constitutive-eq-model} are nothing but $\gamma_n v^n$ and $\gamma_t v^a$. Therefore, starting from \autoref{eq:membrane-eq}, \autoref{eq:activators-eq} and \autoref{eq:covariant-conservation}, and using $\partial^a = g^{ab}\partial_b$ with $g^{ab}$ the metric tensor associated with the membrane manifold, we obtain the following hydrodynamic equations for the membrane height $h(\bm{x},t)$ and the activator density $\rho(\bm{x},t)$:
\begin{widetext}\begin{subequations}\label{eq:active-interface-equations}\begin{align}
\label{eq:active-interface-equations-height}\partial_t h(\bm{x},t) &= \frac{\sqrt{g}}{\gamma_n}\left( f(\rho) -\frac{\delta \mathcal{F}[h,\rho]}{\delta h}  + \sqrt{2k_B T_n\gamma_n}\xi_n(\bm{x},t)\right);\\
\label{eq:active-interface-equations-density}\partial_t (\sqrt{g} \rho(\bm{x},t)) &=\frac{1}{\gamma_t} \partial_a\left[ \sqrt{g}g^{ab}\left(\rho\partial_b \frac{\delta \mathcal{F}[h,\rho]}{\delta \rho} + \sqrt{2k_BT_t\gamma_t\rho}\xi_b(\bm{x},t)\right)\right] + \partial_a\left(\frac{g^{ab}}{\sqrt{g}}\rho(\partial_b h)\partial_t h\right).
\end{align}\end{subequations}\end{widetext}

A few comments are in order. First, the height equation~\autoref{eq:active-interface-equations-height} resembles Model A of critical dynamics~\cite{hoenberghalperin-critical} for the relaxation towards equilibrium of a non-conserved order parameter, with an additional constraint of symmetry with respect to shifts of the height $h(\bm{x},t)\to h(\bm{x},t)+c$. The only changes with respect to the conventional model A dynamics are the active force $f(\rho)$ and the  factor $\sqrt{g}$  multiplying the mobility $\gamma_n^{-1}$. The function $f(\rho)$ remains, for now, unspecified, and will be further discussed later. %Its value at $\rho\,{=}\,0$ represents an external driving force which sets the membrane into motion even in the absence of activators, the first derivative $f'(\rho)|_{\rho=0}$ equals the force imparted on the membrane per single activator. \mre{ We now turn to the density equation  (\ref{eq:active-interface-equations-density}).} 
Second, the first term on the right-hand side of the density equation, (\ref{eq:active-interface-equations-density}), resembles model B of critical dynamics~\cite{hoenberghalperin-critical}, which describes the relaxation of a globally conserved order parameter. The factors $\sqrt{g}$ in the time-derivative and $\sqrt{g}g^{ab}$ in the square-bracket term are required because the density is defined on a curved manifold rather than in flat space. In addition, the final term in the right hand side of~\autoref{eq:active-interface-equations-density}, proportional to $\partial_t h$, represents the non-dissipative kinematic coupling between activator density and mem
brane slope generated by the membrane motion~\cite{cai1995}, which will play a key role in what follows.

Although ~\autoref{eq:active-interface-equations-density} describes a collection of point-like activators, hard excluded volume interactions can be implemented by replacing the factors of $\rho$ multiplying the free energy derivative and the kinetic current and that appearing in the noise coefficient with $\rho\left(1-\rho\right)$. Similar terms can be obtained by considering soft interactions via an extra term in the free energy proportional to the squared density. Additional deterministic terms In~\autoref{eq:active-interface-equations-density}'s right-hand side might also result from the inclusion of tangential active forces~\cite{chen2020, alimohamadi2020non}. Finally, let us remark that the model equations can be readily extended to describe several possible species of activators, each having its own density and a distinct interaction with the membrane.

\subsection{Hydrodynamic equations for small density and height fluctuations}

In preparation for our field-theoretical analysis, we now expand  \autoref{eq:active-interface-equations} around the homogeneous solution where the activators are evenly spread across the membrane. As, in this solution, $ \rho(x,t) = \rho_0 = $ constant, the active force acts as a passive homogeneous driving force $f(\rho_0)$ in this case, giving rise to an overall movement of the membrane with velocity $\lambda=f(\rho_0)/\gamma_n$. Small fluctuations about the homogeneous solution are accounted for by setting $\rho = \rho_0 + \phi$ and $h = \lambda t + \delta h$. Consistently with the Monge gauge prescription, we will also assume the interface slopes to be small, i.e. $(\partial_a \delta h)^2 \ll 1$. Choosing to expand the equations around the flat homogeneous solution is convenient for the nearly-flat phase analysed in this paper. However, we shall point out that the interaction between membrane and proteins leads to a rich phase diagram of possible stationary shapes, both in the  presence and absence of active forces~\cite{kabaso2011theoretical, fosnaric2019theoretical, sadhu2021modelling}.

To write down the equations of motion it is also necessary to specify a functional form for the free energy $\mathcal{F}[h,\rho]$. For the membrane contribution, we consider a simple surface tension term, written in covariant form as $\nu\int d^dx \sqrt{g}$, with $\nu$ the surface tension. 
For the density-dependent part we consider the ideal gas entropy, written in covariant form as $k_BT_t \int d^dx\sqrt{g} \rho \log{\rho}$. These two contributions give $\mathcal{F}_{\text{free}}$, the free energy without interactions, which becomes, in terms of the height fluctuations $\delta h = h(\bm{x},t)\,{-}\,\lambda t$ and the excess density $\phi=\rho(\bm{x},t)-\rho_0$, 
\begin{equation}\label{eq:free-free-energy}
\mathcal{F}_{\text{free}} = \frac{\nu}{2}\int d^dx\, (\partial_a \delta h)^2 +  \frac{k_BT}{2\rho_0} \int d^dx\, \phi^2  + \mathcal{O}(\partial h^4, \phi^4).
\end{equation}

We now add the contribution to the free energy from interactions. Following classical studies on the relation between the membrane shape and composition~\cite{markin1981lateral}, active membrane theories have typically considered a coupling between the activator density and the membrane curvature~\cite{ramaswamy2000,ramaswamy2001, goutaland2020}. This coupling is inspired by the fact that most of the proteins responsible for activating the membrane dynamics can be bound to a particular banana-shaped dimer---known as the BAR domain---which causes them to acquire an intrinsic curvature~\cite{habermann2004bar,vogel2006local}. The intrinsic curvature of the activators interacts with the membrane curvature, so that the free energy is minimised when activators sit in regions of the membranes with local curvature close to their intrinsic curvature.
While the free energy contribution of the BAR domains should also take into account their strongly anisotropic shape~\cite{mesarec2016closed}, we model this interaction with the following simplified term, valid for our isotropic, point-like activators,
\begin{equation}\label{eq:int-free-energy}
\mathcal{F}_{\text{curv}} = -c \int d^dx\, \left[(\partial_a^2 \delta h)(\rho_0 + \phi)\right]. 
\end{equation}
The free energy above can be derived from a microscopic description of a curved inclusion embedded within the lipid bilayer, in the limiting case where all the lipids are aligned to the membrane normal~\cite{helfrich1988intrinsic, kralj1999free, may2000protein, fosnaric2006influence}. In this case the constant $C$ is proportional to the activators' intrinsic curvature. The sign of $c$ determines whether the intrinsic curvature of the activators is positive ($c\,{>}\,0$) or negative ($c\,{<}\,0$). Expanding all the other terms of \autoref{eq:active-interface-equations} in powers of $\delta h$ and $\phi$,  we obtain our model equations for excess protein density and fluctuating height.

To simplify the notation,  from now on we will replace the symbol $\delta h$ with $h$. The resulting equations read
\begin{subequations}\label{eq:active-interface-fluctuating-feterms}
\begin{align}\label{eq:a-i-f-int-fet}
\partial_t h &=  a_h\phi + \frac{\alpha}{2}\phi^2 -c_h \partial_a^2 \phi + \nu_h \partial_a^2 h +\frac{\lambda}{2}\left(\partial_a h\right)^2 \nonumber\\ & + \sqrt{2 D_h}\xi_n,\\ 
\label{eq:a-i-f-dens-fet}
\partial_t \phi &= a_\phi\partial_a^2 h - c_\phi \partial_a^4 h + \nu_\phi \partial_a^2 \phi+ \lambda\partial_a\left(\phi\partial_a h\right) \nonumber\\ &+ \partial_a\left(\sqrt{2 D_\phi}\xi_a\right).
\end{align}
\end{subequations}
%\begin{subequations}\label{eq:active-interface-fluctuating}
%\begin{align}\label{eq:a-i-f-int}
%\partial_t h &=  a_h\phi  -c_h \partial_a^2 \phi + \nu_h \partial_a^2 h +\frac{\lambda}{2}\left(\partial_a h\right)^2+ \nonumber\\ & + \sqrt{2 D_h}\xi_n,\\ 
%\label{eq:a-i-f-dens}
%\partial_t \phi &= a_\phi\partial_a^2 h - c_\phi \partial_a^4 h + \nu_\phi \partial_a^2 \phi+ \lambda\partial_a\left(\phi\partial_a h\right)+ \nonumber\\ &+ \partial_a\left(\sqrt{2 D_\phi}\xi_a\right).
%\end{align}
%\end{subequations}
Regarding the physical interpretation of parameters, we have already discussed two fundamental ones: the speed of the membrane $\lambda\,{=}\,f(\rho_0)/\gamma_t$ and average density of activators $\rho_0$. Two other parameters  appearing in  \autoref{eq:active-interface-fluctuating-feterms}  are $a_\phi = \rho_0 \lambda$, measuring the advection of density fluctuations by the membrane slopes, and $a_h = f'(\rho_0)/\gamma_n$, which quantifies the additional membrane speed due to fluctuations in the density profile. In addition, we have $\nu_h = \nu/\gamma_n$, $\nu_\phi = k_bT_t/\gamma_t$, $c_h=-c/\gamma_n$ and $c_\phi = \rho_0 c /\gamma_t$. 
%The noise coefficients $D_h$ and $D_\phi$ should, in principle, be chosen so as to satisfy the fluctuation-dissipation theorem. 
Given that the system is driven out of thermodynamic equilibrium by the active force $f(\rho)$ acting on the membrane, $D_h$ and $D_\phi$ need not satisfy the fluctuation-dissipation theorem and instead can be considered as independent parameters, encoding both thermal fluctuations and those caused by energy-consuming processes. %Eq.~(\ref{eq:active-interface-fluctuating}) represents a baseline theory of membrane-activator dynamics, which keeps only the lowest order terms. 
The term $\alpha \phi^2$ in \eqref{eq:a-i-f-int-fet}  comes from   the second-order expansion of $f(\rho)$ about $\rho_0$, with $\alpha=f''(\rho_0)/\gamma_n$. This term quantifies many-body effects~\cite{zakine2018} in activator-mediated interface growth, and would arise physically, for instance, in the case where activators are dilute ($\rho_0\ll 1$) and stimulate growth only when they dimerise.

However, \eqref{eq:active-interface-fluctuating-feterms} is not yet  complete.
Importantly, we also need to consider terms which  cannot be written as $\partial_a \mu_{eq}$ with $\mu_{eq}$ the ``equilibrium chemical potential'' $\delta\mathcal{F}/\delta \rho$. We require two additional terms  representing the gradient of a nonequilibrium current $j^{neq}_a$.  One part of $j^{neq}_a$ can  be written as the gradient of a ``nonequilibrium chemical potential'' proportional to the squared slope $(\partial_a h)^2$ considered before in theories of conserved kinetic roughening~(see chapter 5 of \cite{krug1997origins} and reference therein), whereas another part is proportional to the Laplacian of $h$ times its gradient (this term has been considered recently~\cite{Caballero18,Skultety2021}, again in the context of conserved surface roughening). Including these terms, our full set of equations reads
\begin{subequations}\label{eq:active-interface-fluctuating-full}
\begin{align}\label{eq:a-i-f-int-full}
\partial_t h &=  a_h\phi + \frac{\alpha}{2}\phi^2 -c_h \partial_a^2 \phi + \nu_h \partial_a^2 h +\frac{\lambda}{2}\left(\partial_a h\right)^2 \nonumber\\ & + \sqrt{2 D_h}\xi_n,\\ 
\label{eq:a-i-f-dens-full}
\partial_t \phi &= a_\phi\partial_a^2 h - c_\phi \partial_a^4 h + \nu_\phi \partial_a^2 \phi+ \lambda\partial_a\left(\phi\partial_a h\right) \nonumber\\ &+\frac{\kappa}{2}\left[\partial_a^2 (\partial_b h)^2 -2\partial_a\left((\partial_a h)\partial_b^2 h\right)\right] \nonumber \\ &+ \partial_a\left(\sqrt{2 D_\phi}\xi_a\right).
\end{align}
\end{subequations}
The coefficient $\kappa$ of the additional nonlinear terms in \eqref{eq:a-i-f-dens-full} cannot be related to any of the other parameters of the problem since the corresponding term cannot be derived within the framework of \autoref{eq:active-interface-equations}. 
As we later demonstrate by analysing the primitive degree of divergence of perturbative corrections, any additional nonlinear term in~\autoref{eq:active-interface-fluctuating-full}---such as those coming from local interactions between activators---would be irrelevant in the RG sense, irrespective of the scaling regime considered. %In fact some of the terms in~\autoref{eq:active-interface-fluctuating-full} turn out to be irrelevant for  specific scaling regimes.
 %The inclusion of $\kappa$ in our theory is required for renormalization purposes, as the corresponding vertex function has a positive primitive degree of divergence (cf.~\autoref{sec:renormalization-diffusive} and~\autoref{sec:renormalization-diffusive-2}).}

We end this section with a technical note. By construction, \autoref{eq:active-interface-equations} is invariant with respect to reparametrisation of the membrane within the Monge gauge. Additionally, also transformations of the external, $(d+1)$-dimensional, coordinate system should leave \autoref{eq:active-interface-equations} invariant, provided they do not break the structure of the Monge gauge. Among these transformations there are infinitesimal membrane tilts~\cite{KPZ}. When the equations of motion are expanded about the homogeneous solution $h(\bm{x},t)=\lambda t$, $\rho(\bm{x},t) = \rho_0$, this symmetry is equivalent to the following set of transformations: 
\begin{equation}\label{eq:tilt-transformation}
\begin{array}{c} \bm{x} \\ h(\bm{x}) \end{array} \to \begin{array}{ccc} \bm{x}' &=&  \bm{x}-\bm{\epsilon}\lambda t \\h'(\bm{x}',t)  &=& h(\bm{x},t) + \bm{\epsilon}\cdot \bm{x}\end{array}.
\end{equation}
Care must be taken that the additional nonlinear nonequilibrium terms, whose inclusion 
we have argued  is required by the RG method, satisfy this symmetry---that is how the relative weight of the two contributions to the $\kappa$-term in \eqref{eq:a-i-f-dens-full}  are fixed. Additionally, the symmetry implies the coincidence of certain perturbative corrections which can be exploited to simplify the RG calculations.

\subsection{Field-theoretic formalism}\label{ssec:formalism}

Before proceeding with the analysis of the scale-invariant behavior of the model summarised by \autoref{eq:active-interface-fluctuating-full}, it is convenient to introduce an equivalent formulation of the model, based on the functional probability of the fields $h$ and $\phi$ rather than on the stochastic equations of motion. The path probability can be obtained from that of the noises $\xi_n$ and $\xi_a$, which is Gaussian, following the procedure used by Onsager and Machlup for the linear Langevin equation~\cite{onsager-machlup1,onsager-machlup2}. The resulting probability can then be written in a simpler form, at the price of introducing one auxiliary field for every field in the theory. This procedure is commonly credited to Martin, Siggia and Rose~\cite{martin-siggia-rose}, De Dominicis~\cite{dedominicis} and Janssen~\cite{janssen}---the details can be found in critical dynamics textbooks such as~\cite{tauberCriticalDynamics,Vasilev04}. The result is the path probability of the fields $P[h,\phi]$ written in the following form,
\begin{equation}\label{eq:path-probability}
 P[h,\phi] = \int \mathcal{D}[i\ti{h}]\mathcal{D}[i\ti{\phi}]e^{-S[\ti{h},h,\ti{\phi},\phi]},
\end{equation}
where the action $S[\ti{h},h,\ti{\phi},\phi]$ is given by
\begin{widetext}\begin{equation}\label{eq:action}
\begin{aligned}
 S[\ti{h},h,\ti{\phi},\phi] = \int d^d\bm{x}dt\, &\left[\ti{h}\left(\partial_t - \nu_h \nabla^2 \right)h + \tilde{\phi}\left(\partial_t -\nu_\phi \nabla^2 \right)\phi \right. \\ &\left. - D_h\ti{h}^2 - D_\phi\left(\nabla \ti{\phi}\right)^2 - \ti{h}\left(a_h -c_h\nabla^2\right)\phi -\tilde{\phi}\left(a_\phi - c_\phi \nabla^2\right)\nabla^2h \right. \\ &\left. -\frac{\lambda}{2}\ti{h}\left(\nabla h\right)^2 + \lambda \left(\bm{\nabla}\ti{\phi}\right)\cdot \left(\phi\bm{\nabla}h\right)\right. \\ & \left.-\frac{\alpha}{2}\ti{h}\phi^2 +\frac{\kappa}{2}\left(\bm{\nabla}\ti{\phi}\right)\cdot\left( \bm{\nabla} (\nabla h)^2 - 2(\bm{\nabla}h) \nabla^2 h \right)\right].
\end{aligned}
\end{equation}\end{widetext}
In the definition above, the first two lines on the right-hand side correspond to the linearised equation and the last two to the %lowest-order 
nonlinear terms. It is worth remarking that \autoref{eq:path-probability}, with the action from \autoref{eq:action}, is completely equivalent to the stochastic PDEs formulation of \autoref{eq:active-interface-fluctuating-full}.

The harmonic, or Gaussian, part of the action, containing contributions which are at most quadratic in the fields, corresponds to the linearised stochastic equations. It can be written compactly in Fourier space as (with $\Bbbk=(\bm{k},\omega)$ and $\bm{\psi}=(\ti{h},h,\ti{\phi},\phi)$)
\begin{equation}\label{eq:gaussian-action}
 S_0[\bm{\psi}] = \frac{1}{2}\int_{\Bbbk} \bm{\psi}(-\Bbbk)\cdot\left[\bm{A}_0(\Bbbk)\bm{\psi}(\Bbbk)\right].
\end{equation}
The linear coupling matrix $\bm{A}_0(\Bbbk)$ is explicitly given in~Appendix A. %\autoref{app:feynman}. 
The inverse of this matrix yields the correlations of the Gaussian model via
\begin{equation}\begin{aligned}\label{eq:gaussian-correlations}
 \avg{\psi_i(\Bbbk_1)\psi_j(\Bbbk_2)}_0 &= \left(A_0^{-1}\right)_{ij} \delta(\Bbbk_1+\Bbbk_2) \\
 &= C_0^{\psi_i\psi_j}(\Bbbk_1)\delta(\Bbbk_1+\Bbbk_2),
\end{aligned}\end{equation}
where $\delta(\Bbbk_1+\Bbbk_2)$ is a shorthand for $(2\pi)^{d+1}\delta(\bm{k}_1+\bm{k}_2)\delta(\omega_1+\omega_2)$. \autoref{eq:gaussian-correlations} can be derived by interpreting $e^{-S[\bm{\psi}]}$ as the joint path probability of physical ($h$ and $\phi$) and response ($\ti{h}$ and $\ti{\phi}$) fields respectively -- i.e., $P[\bm{\psi}]=e^{-S[\bm{\psi}]}$. Introducing a conjugate current for each of the fields, $\bm{j}=(j_h,j_{\ti{h}},j_\psi,j_{\ti{\psi}})$, and averaging $e^{\int d^dx dt\, \bm{j}\cdot\bm{\psi}}$ over $P[\bm{\psi}]$ yields the moment generating functional of the fields,
\begin{equation}\label{eq:moment-gen-fun}
 Z[\bm{j}] = \int \mathcal{D}[\bm{\psi}] e^{\int d^dx dt\, \bm{j}\cdot\bm{\psi}}e^{-S[\bm{\psi}]}.
\end{equation}

If the action is quadratic in the fields, the integral on the right-hand side of \autoref{eq:moment-gen-fun} is a simple Gaussian integral. \autoref{eq:gaussian-correlations} then follows by differentiating the result of the integral twice with respect to the components of $\bm{j}$, and then setting $\bm{j}\to 0$. While $Z[\bm{j}]$ generates the moments of the fields -- i.e., the $n$-point correlation functions -- its logarithm $\ln{Z[\bm{j}]}$ generates the cumulants of the fields -- i.e., the $n$-point {\it connected} correlation functions. The Legendre-Fenchel transform of $\ln{Z[\bm{j}]}$, 
\begin{equation}\label{eq:vertex-gen-fun}
 \Gamma[\bm{\psi}] = \sup_{\bm{j}} \left\lbrace \int d^dx dt\, \bm{\psi}\cdot\bm{j} - \ln{Z[\bm{j}]}\right\rbrace,
\end{equation}
is the generating functional of $n$-point {\it vertex functions}, usually referred to as the {\it effective action}~\cite{amitFieldTheory,Vasilev04}. %~\footnote{$\bm{\psi}$ here are technically the `classical' fields (expectations of the fields w.r.t. the path probability), but we will use the same notation for convenience.}
In general, its relation to the action functional is the following \cite{amitFieldTheory,Vasilev04}
\begin{align} \label{eq:effective-action}
	\Gamma[\bm{\psi}] \,{=}\,S[\bm{\psi}]+\text{(loop corrections)}.
\end{align}
For an action which is at most quadratic in the fields, the corrections are absent and $\Gamma[\bm{\psi}]$ coincides with the action itself. The $n$-point vertex functions can thus be thought of as the coefficients of the expansion of the action in powers of the fields. For general nonlinear models, perturbative corrections must be taken into account, which can be computed from the perturbative corrections to correlation functions (see~Appendix B%\autoref{app:perturbationKPZ}
,~\autoref{eq:correlations-to-vertices}).

In each of the regimes that will be considered in the remainder of the paper, a simple power counting analysis allows the identification of the {\it upper critical dimension} $d_c$ of the nonlinear coupling parameters---the coefficients of the nonlinear terms in the equations of motion. Above $d_c$, all the nonlinearities in the action are irrelevant from the perturbative point of view, and the scale-invariant properties of the model are the same as those of the linearised counterpart. In other words, {\it mean-field} scaling is exact above $d_c$. In contrast, below $d_c$ the mean-field scaling exponents gain non-trivial corrections due to the nonlinearities of the model. Once $d_c$ is identified, the terms in the action~\autoref{eq:action} are split into three classes: {\it relevant} terms, {\it marginal} terms and {\it irrelevant} term. The relevant terms are those which drive the system away from criticality: they must be set to zero in order to find scale-invariant behavior, otherwise the RG flow takes the system to a different scaling regime. The irrelevant terms do not influence the critical properties of the model and the marginal terms are those that actually need to be accounted for. Close to the upper critical dimension, where it is reasonable to assume that the coefficients of the nonlinear couplings are actually small, the effects of the nonlinearities can be examined with perturbation theory, the details of which are given in~Appendix A. In particular, the marginal terms of the action acquire perturbative corrections which are generally expressed as momentum integrals. These momentum integrals might diverge at the upper critical dimension because of the large-$k$ properties of the integrand-- i.e., they show ultraviolet (UV) divergences.

The aim of the remormalization procedure is that of absorbing the UV divergences of perturbative corrections into a finite set of  {\it renormalized} coefficents (and fields)---the details of the procedure are given in~\autoref{sec:renormalization-diffusive} in the context of the active KPZ model. Studying how the renormalized coefficients vary with the scale of observation ultimately allows the derivation of flow equations and the identification of the fixed (scale invariant) points of these equations. The latter represent different possible universality classes for the model, to which the model flows (in the renormalization group sense) depending on the starting value of the original parameters. We now proceed with a detailed analysis of the scale-invariant behavior of the active interface model defined by \autoref{eq:active-interface-fluctuating-full}.

\section{Mean-field theory of active membranes}\label{sec:two}

In the present section we will analyze the properties of the model summarised in~\autoref{eq:active-interface-fluctuating-full} at the mean-field level. The first step is to determine the conditions under which the system is linearly stable. This will provide information about different scaling regimes accessible within our approach, as well as representing the starting point for beyond-mean-field calculations.

\subsection{Linear stability}\label{sec:linstab}

\begin{figure}[t!]
	\centering
	\includegraphics[width=0.5\textwidth]{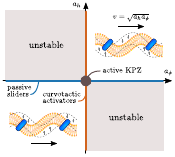}
	\caption{Phase diagram of the active interface model in the $a_h-a_\phi$ plane. The linearly-unstable (second and fourth quadrant) cannot be accessed within our framework. All membranes with strictly positive $a_h a_\phi$ can be characterised with the same set of scaling exponents, to which we refer as the ``generic active membrane'' universality class. As shown by the cartoon, the dynamics in this region of the parameter space is dominated by coupled density and height waves having speed $v\propto\sqrt{a_h a_\phi}$~\cite{ramaswamy2000, cagnetta1}. Tuning the wave speed to zero leads to new scaling regimes whose properties are discussed in the text.}
	\label{fig:phase-diagram}
\end{figure}

The linearised, noiseless version of~\autoref{eq:active-interface-fluctuating-full} reads
\begin{subequations}\label{eq:active-interface-fluctuating-lin}
	\begin{align}\label{eq:a-i-f-int-lin}
	\partial_t h &=  a_h\phi  -c_h \partial_a^2 \phi + \nu_h \partial_a^2 h\\
	\label{eq:a-i-f-dens-lin}
	\partial_t \phi &= a_\phi\partial_a^2 h - c_\phi \partial_a^4 h + \nu_\phi \partial_a^2 \phi\;.
	\end{align}
\end{subequations}
A Fourier-space solution of the linearised equations of motion, with $h(\bm{x},t), \phi(\bm{x},t) \propto e^{i \left({\bm{k}}\cdot {\bm{x}}-\omega t\right)}$, yields the following conditions,
\begin{subequations}\label{eq:stability-condition}
	\begin{align}
	\nu_h + \nu_\phi > 0 , \label{eq:stability-conditiona}\\
	\Delta\equiv\nu_h \nu_\phi k^4 +(a_h + c_h k^2)(a_\phi k^2+ c_\phi k ^4)>0,\label{eq:stability-conditionb}
	\end{align}
\end{subequations}
for the linear stability of the $k$-th height and density modes with respect to perturbations in the height and density profiles. The dispersion relation, linking mode frequency and wavevector, is given by
\begin{equation}\label{eq:dispersion-relation}from now one
i\omega = \frac{1}{2}\left(\left(\nu_h+\nu_\phi\right)k^2 \pm \sqrt{\left(\nu_h+\nu_\phi\right)^2k^4-4\Delta}\right).
\end{equation}
The first stability condition  \eqref{eq:stability-conditiona}  is satisfied in the physically relevant case
$\nu_h >0$ and $ \nu_\phi>0$. The second condition shows that there is an infrared ($k\to 0$) instability for $a_h a_\phi <0$, which can be achieved when $f(\rho_0)$ and $f'(\rho_0)$ have opposite signs. This situation can be realised, for instance, with $f(\rho) = a - b \rho$, where $a$ and $b$ have the same sign and $a-b\rho_0>0$. Such a functional form of $f(\rho)$ might be suitable to describe a membrane driven by a homogeneous force $a$ and an active force linearly proportional to the activator density, $b\rho$, which act in opposition to each other. If the instability persists at the nonlinear level, the assumption of vanishing slopes $(\partial_a h)^2\ll 1$ might break down together with the Monge-gauge description. This would force us to consider a more generic membrane parametrisation $\mathbb{X}(\bm{x})$, as  is done for instance in~\cite{kabaso2011theoretical, fosnaric2019theoretical, sadhu2021modelling}. Therefore, from now on we restrict our analysis to the $a_h a_\phi \geq 0$ portion of the parameter space in Fig.~\ref{fig:phase-diagram}. 

Another instability arises from  condition \autoref{eq:stability-conditionb} at intermediate wavevectors if $\nu_h \nu_\phi + a_h c_\phi + c_h a_\phi<0$. This type of instability was discussed in~\cite{gov2006}, and arises because of the mixing between curvature coupling and active growth. For instance, if $a_h\,{>}\,0$,
activators with negative intrinsic curvature, i.e. $c_\phi\,{<}\,0$, tend to cluster due to a positive feedback: localised growth causes a bump with negative curvature, which recruits more activators due to the curvature coupling, which in turn cause a larger localised growth. This mechanism leads to an instability when its strength $a_h c_\phi\,{=}\,-a_h|c_\phi|$ exceeds that of the combination of the activators' diffusion and surface relaxation $\nu_h \nu_\phi$. Analogously, activators with positive intrinsic curvature cluster when $a_h\,{>}\,0$.
Finally, an ultraviolet instability ($k\to\infty$) occurs when $c_\phi c_h\,{<}\,0$, as is the case since $c_h\,{=}\, -c/\gamma_n$ and $c_\phi\,{=}\, \rho_0c/\gamma_t$. The latter instability, which can be easily cured by adding a bending rigidity term to the membrane free-energy, does not influence the result of this paper, as in the regimes to be considered $c_h$ and $c_\phi$ do not simultaneously appear in the equations of motion.

In the cases where $\Delta\,{>}\,0$ , such that there is no linear instability in the model, the nature of the dispersion relation close to $k\to 0$ depends on the value of $a_h a_\phi$. For $a_h a_\phi\ne 0$ we find an \emph{acoustic} dispersion law, $\omega \simeq \pm (a_h a_\phi)^{1/2} k$, whereas for $a_h a_\phi =0$ we find a \emph{diffusive} law, $ \omega \sim k^2$. This points to different dynamical scaling regimes in the two cases, to be discussed below, which will have important consequences for the critical properties of the system. Even within the diffusive regime, the other scaling properties depend on how the limit $a_h a_\phi=0$ is approached: either one of the parameters $ a_{h} $ or $ a_{\phi} $ is being set to zero while the other one is kept finite, or both go to zero at the same time. Each of these three cases represents a different instance of the diffusive regime, with, as will be shown below, different scaling exponents and upper critical dimension. In total, four classes can be identified:
\begin{itemize}
	\item Acoustic scaling ($ \omega \sim k $ in mean-field)
	\begin{itemize}
		\item Generic Active Membrane--for $ a_{h}a_{\phi} > 0 $
	\end{itemize} 
	\item Diffusive scaling ($ \omega \sim k^{2} $ in mean-field)
	\begin{itemize}
		\item Active KPZ--for $ a_{h}=a_{\phi} = 0 $
		\item Curvotactic Activators--for $ a_{h} = 0, \ a_{\phi} > 0 $
		\item Passive Sliders--for $ a_{h} > 0, \ a_{\phi} = 0 $
	\end{itemize}
\end{itemize}
These scaling regimes are summarized in the phase diagram of~\autoref{fig:phase-diagram}

Before proceeding with the mean-field study of all the above classes, let us discuss in more detail the relation of  the linear system \autoref{eq:active-interface-fluctuating-lin} with other active membrane models in the nearly-flat phase. Similar equations, for instance, arise when the membrane is driven by two families of ``active pumps'', with intrinsic curvatures of opposite signs and pushing the membrane in opposite directions. This model was studied in~\cite{ramaswamy2000,ramaswamy2001,maitra2014}. In such a model, $\rho$ becomes the {\it difference} in local density of pumps of different kinds, rather than the absolute density of activators. The analysis of~\cite{ramaswamy2000}, which is mostly concerned with the case where the average density of pumps of the two kinds coincide (corresponding to $a_\phi=0$ in \autoref{eq:active-interface-fluctuating-full}), shows the onset of an instability driven by the coupling with the curvature. Because of this instability, pumps of the same kind cluster and the membrane develops finger-like protrusions. When, instead, one species of pumps exceeds the other, perturbations in the signed density profile travel as waves along the membrane\cite{ramaswamy2000,maitra2014}.

Subsequent works in the active membrane literature have focussed on the interplay between active forces and curvature coupling. For instance, the model studied in~\cite{gov2006},  which also leads to the intermediate wavevector instability in \autoref{eq:stability-condition}, can be obtained from \autoref{eq:active-interface-fluctuating-lin} by setting $a_\phi$ and $c_h$ to zero, while having non-vanishing $a_h$ and $c_\phi$. 
(However, we note the limiting case $a_\phi, c_h\to 0$ differs from the equations in~\cite{gov2006} by a hydrodynamic interaction kernel in the membrane dynamics.) This model describes diffusing activators which have a preference for a certain sign of membrane curvature, but importantly  does not include a coupling between activator density and interface slope. It shows  unstable or wave-like behavior, depending on the specific value of the curvature coupling. In particular, in the unstable phase, the curvature coupling contributes to the phase separation of the activators~\cite{veksler2007}.

\subsection{Power counting: acoustic and diffusive scaling}\label{sec:powercounting}

We now begin our renormalization group (RG) study with a power counting analysis, which allows us to compute the {\it engineering dimensions} of fields and parameters. This procedure is a useful preliminary analysis prior to a full RG calculation: the engineering dimensions describe the critical behavior, or scaling, of the model when the effects of nonlinear terms are neglected. Engineering dimensions are computed by introducing an arbitrary momentum scale $\mu$ such that
\begin{equation}\label{eq:eng-dim}
 \sdim{x} = \mu^{-1},\, \sdim{t}=\mu^{-z},\, \sdim{\psi} = \mu^{y_\psi}, 
\end{equation}
where $\sdim{.}$ denotes the dimensionality of the object between brackets and $\psi$ denotes a generic parameter or field. The quantity $z$ is the {\it dynamic exponent}, which specifies the relation between timescales and lengthscales in the thermodynamic limit $x,t\to \infty$ (or $k,\omega\to 0$). Each term of the action of \autoref{eq:action} has a specific dimensionality which can be obtained by combining the engineering dimensions of fields, parameters, derivatives and differentials. The action itself, however, is dimensionless by definition and so should be each of the terms it contains. Therefore, every term of the action yields an algebraic relation between the engineering dimensions which must be satisfied in order to render that term dimensionless. For instance, the term $\int d^dx dt\, \ti{h}\partial_t h$ gives the equation $y_h+y_{\ti{h}}-d\,{=}\,0$.

The inversion of all these algebraic relations yields the engineering dimensions of all fields and parameters. We begin by inverting the algebraic relations stemming from the harmonic part of the action, thus determining the mean-field scaling of the model. The harmonic action contains 10 terms, but 12 fields and parameters. Therefore, the system of equations that determines the corresponding engineering dimensions appears at first sight underdetermined. Some physical considerations come to the rescue. First, the mean-field value of the dynamical exponent $z$ can be computed by analysing the dispersion relation of the linearised model, \autoref{eq:dispersion-relation}. Thus, for $a_h a_\phi>0$, we have  acoustic behavior, with $ \omega \sim k$ for $k\to 0$, so that $z=1$, whereas for $a_h a_\phi=0$ we have  diffusive behavior, $\omega \sim k^2$ for $k\to 0$, so that $z=2$.

Second, we note that the stochastic terms in the equations of motion~\autoref{eq:active-interface-fluctuating-full} can be included in scale invariant descriptions of the system only if they are marginal. Otherwise, they either can be dropped (if irrelevant) or should be tuned to $0$ to yield a scale-invariant description (if relevant). As we are interested in cases where stochastic effects are present, so that it is meaningful to speak about height and density correlations, we require in what follows that at least one of the stochastic terms is marginal. Because the two equations are coupled, assuming a non-zero value of the noise parameter in one equation transfers stochastic effects to the other equation if the appropriate coupling parameter is non-zero. For instance, if $a_\phi\ne 0$, noise in the equation for $h$ causes stochastic effects in the equation for $\phi$, even if the latter does not itself include a noise term.
%We therefore assume in what follows that for each case to be considered at least one of these is marginal: this assumption is then validated {\it a posteriori} with a consistency check on the resulting set of exponents, as we shall discuss. 
In the following discussion, we begin by assuming both noise terms to be marginal, 
%Since the squared-response-fields terms of the action correspond to the stochastic terms of the equations of motion~\autoref{eq:active-interface-fluctuating},  we require  $D_h \ti{h}^2$ and  $D_\phi(\nabla \ti{\phi})^2$ to be marginal that is}
\begin{equation}\label{eq:noise-assumption}
 y_{D_h}=y_{D_\phi}=0,
\end{equation}
and later on show that to encompass all possible scenarios we also need to consider the cases where only one of the noise terms is marginal. 
%Since the squared-response-fields terms of the action correspond to the stochastic terms of the equations of motion~\autoref{eq:active-interface-fluctuating}, the purpose of \autoref{eq:noise-assumption} is that of keeping the stochastic terms marginal in the scaling limit. \mre{The foll discussion could be removed: In fact, if all the stochastic terms were found to be irrelevant in the scaling limit, the model would be effectively described by deterministic equations of motion and the construction of the action functional would break down.}{\color{red} [Technically the response functional still exists, but you can integrate over the response field and get the deterministic equations of motion]} 
% Note that condition \autoref{eq:noise-assumption} (or the weaker condition of $ y_{D_h}=0$ or $y_{D_\phi}=0$) is equivalent to considering a set of equations where we have performed a suitable rescaling of the fluctuating fields, which we will detail in each case. 

Once the engineering dimensions of fields and linear couplings are extracted from the harmonic part of the action, nonlinear terms dictate the scaling dimensions of the coupling constants $ \lambda,\alpha $, and $ \kappa $, which in turn determine their `naive' upper critical dimension. We now consider the engineering dimensions obtained within the acoustic ($z\,{=}\,1$) and diffusive ($z\,{=}\,2$) scaling separately and comment on their implications for the static properties of active membranes.

\subsection{Scaling laws in the acoustic regime: the generic active membrane model}\label{ssec:acoustic scaling}

The acoustic regime corresponds to a dynamic exponent $z=1$, which is physically realised when $a_h a_\phi>0$. This corresponds to the generic active membrane model in Fig.~\ref{fig:phase-diagram}, and is the typical situation for positive $a_h$ and $a_\phi$, which corresponds to the biophysically relevant case of an advancing cellular membrane where activators favour membrane growth. Having set $z=1$ and $y_{D_h}=y_{D_\phi}=0$, we can find the engineering dimensions for all other fields in this acoustic regime. The values are summarised in Table~\ref{tab:e-d-euler}.
\begin{table}[ht!]
	\def\arraystretch{1.3}
	\centering
	\begin{tabular}{| c || c | c | c | c | c | c | c | c | c |}
		\hline \hline
		$\psi$ & $\ti{h}$ & $h$ & $\ti{\phi}$ & $\phi$ & $\nu_h,\nu_\phi$  & $D_h,D_\phi$ & $a_h,a_\phi$ & $c_h,c_\phi$ & $\lambda$ 
		\\  \hline
		$y_\psi$ & $\frac{d+1}{2}$ & $\frac{d-1}{2}$ & $\frac{d-1}{2}$ & $\frac{d+1}{2}$ & $ -1 $ & $ 0 $ & $0$ & $-2$ & $-\frac{1+d}{2}$
		\\ \hline 
	\end{tabular}
	\caption{Engineering dimensions in the acoustic regime ($z\,{=}\,1$, for $a_h a_\phi>0$, corresponding to the generic active membrane case in Fig.~\ref{fig:phase-diagram}). The other nonlinear couplings $\alpha$ and $\kappa$, like $\lambda$, are irrelevant. Therefore, the engineering dimensions coincide with the true scaling dimensions of the fields for this regime.}.
	\label{tab:e-d-euler}
\end{table}

Notably, the engineering dimensions of the nonlinear coupling parameters, such as $\lambda$, are negative for all positive $d$. Therefore, the mean-field scaling provided by the engineering dimension is exact in all dimensions. This result is useful, as it means that mean field theory is sufficient to describe the generic active membrane model (second and fourth quadrant in the phase diagram in Fig.~\ref{fig:phase-diagram}).% We also find that $a_h$ and $a_\phi$ have zero engineering dimension, so that they need to be retained in the equations of motion describing the scale-invariant system. This is sensible, as in the acoustic regime $a_h a_\phi >0$, so that both terms are non-zero in general.
Therefore, in the acoustic regime, the only couplings which need to be retained in~\autoref{eq:active-interface-fluctuating-full} are $a_h$, $a_\phi$, $D_h$ and $D_\phi$---all other couplings are irrelevant. In order to reduce the number of effective couplings further, we perform the following rescaling of the height and density fields, 
\begin{equation}\label{eq:rescaling-acoustic}
h\to \sqrt{\frac{2D_\phi}{a_\phi^2}} h , \quad \phi \to \sqrt{\frac{2D_h}{a_h^2}} \phi.
\end{equation}
Note that this field rescaling is sensible as both $D_{h,\phi}$ and $a_{h,\phi}$ have zero dimensions, so the numerical factors used to redefine the fields do not change under a momentum (or space) rescaling---i.e., they are scale invariant. Following the rescaling in \autoref{eq:rescaling-acoustic}, the critical properties of the model can be described with the following set of equations in all dimensions:
\begin{subequations}\label{eq:a-i-e-acoustic}
\begin{align}\label{eq:a-i-acoustic-int}
\partial_t h &=  a_\phi\sqrt{D_h/D_\phi}\left(\phi+\xi_n\right),\\ 
\label{eq:a-i-acoustic-dens}
\partial_t \phi &= a_h\sqrt{D_\phi/D_h}\left(\partial_a^2 h+\partial_a\xi_a\right).
\end{align}
\end{subequations}
The noiseless version of these equations has been solved exactly in~\cite{cagnetta3} and the resulting deterministic acoustic waves were studied. With the noise, ~\autoref{eq:a-i-e-acoustic} do not admit a steady state as height and density fluctuations would grow without bound. This can be inferred from the divergence of correlations functions (~\autoref{eq:gaussian-two-point}) for $\nu_h\,{=}\,\nu_\phi\,{=}\,c_h\,{=}\,c_\phi\,{=}\,0$. As a result, these equations are to be interpreted as an inviscid limit $\nu_h, \nu_{\phi}\to 0^+$. Apart from regularising the divergence of steady-state correlations, the presence of infintesimally small viscosities poses some limitations on the range of timescales over which the acoustic scaling holds, as we now discuss.

%\davide{We should note here that, although the diffusive terms with coefficients (or viscosities) $\nu_h$ and $\nu_{\phi}$ are not present  in \autoref{eq:a-i-e-acoustic}, these equations are to  be interpreted as a limit in which $\nu_h, \nu_{\phi}\to 0^+$. Indeed, if the viscous terms were absent altogether,  the system would not  attain a  late-time steady state as height and density fluctuations would grow without bound. One can think of  the viscosities   $\nu_h, \nu_{\phi}\to 0^+$ as being  {\it dangerously irrelevant} parameters. In practice, given very small values of $\nu_h, \nu_{\phi}$,   the acoustic regime scaling which we describe here is observed physically at large scales ($k\sim 1/L$) in a well-defined, but early-time, regime, up to times of order $\sim L$ -- later on, the dangerously irrelevant diffusive terms cause a cross-over to a different dynamical regime, which are not fully characterised within  our  present analysis.}

Since mean-field scaling is exact in the acoustic regime, the scaling dimensions of the fields $h$ and $\phi$ coincide with their engineering dimensions,
\begin{equation}\label{eq:acoustic-scaling}
y_h = \frac{d-1}{2},\quad y_\phi = \frac{d+1}{2}.
\end{equation}
We first note that the dependence of a generic field $\psi$ on $x$ is implied by~\autoref{eq:eng-dim} to be 
\begin{equation}\label{eq:scaling-psi}
\psi \sim x^{-y_\psi}. 
\end{equation}
In Fourier space, with the wavevector $ k = \sqrt{\bm{k}\cdot\bm{k}} $, whose dimension is $[k]=[\mu]$, the scaling dimension of the equal-time height-height correlation $\avg{h(\bm{k},t)h(\bm{q},t)}$ is  $2y_h -2d$ (the term $2y_h$ comes from the two $h$ fields, and the term $-2d$ arises because each Fourier transform entails a $d^dx$ integral which reduces the scaling dimension by $d$). This height-height correlation is related to the time-dependent structure factor of the membrane $S_h(k,t)$ via
\begin{equation}
%\lim_{t\to\infty}\avg{h(\bm{k},t)h(\bm{q},t)} = (2\pi)^2\delta(\bm{k}+\bm{q}) S^h(k).
\avg{h(\bm{k},t)h(\bm{q},t)} = (2\pi)^2\delta(\bm{k}+\bm{q}) S_h(k,t).
\end{equation}
Upon removing the dimension $-d$ of $\delta(\bm{k}+\bm{q})$, the scaling dimension of the time-dependent height structure factor follows:
\begin{equation}\label{height-structure-factor-scaling}
 \sdim{S_h(k,t)} = \mu^{2y_h-d} \Rightarrow S(k,t) \sim k^{2y_h-d} {\cal S}_{h,1}(k t),
\end{equation}
with ${\cal S}_{h,1}$ a scaling function of the dimensionless argument $k^z t$ and the symbol $\sim$ denoting asymptotic equivalence for small wavevectors $k$. Because of the absence of a steady state for $\nu_h = 0 $,  ${\cal S}_{h,1}$ is only well-defined when the system is prepared in a specific initial condition at $t\,{=}\,0$, and it diverges for large times. In fact, concerning systems having large but finite size $L$ with $\nu_h$ begin small but positive, the scaling in~\autoref{height-structure-factor-scaling} is relevant for `early times' $t\sim 1/k$, whereas, after $t\sim(\nu_h k^2)^{-1}$, the system crosses over to a diffusive regime where $S^h(k,t) \sim k^{-2} {\cal S}_{h,2}(k^2t)$. ${\cal S}_{h,2}(k^2t)$ is another scaling function which, unlike ${\cal S}_{h,1}$, converges to a well-defined and time-independent limit as $t\to \infty$. The acoustic behavior at early times influences macroscopic observables such as the time-dependent squared width of the interface, which is obtained as 
\begin{equation}\label{eq:DiscreteToContinuumWidth}
w^2(L,t) = \frac{1}{L^d}\sum_{\bm{k}\neq 0} S_h(k,t).
\end{equation}
When starting from a homogeneous, flat membrane, the early-time acoustic regime leads to width oscillations with period proportional to the system size which are superposed on the usual Edward-Wilkinson growth. Such behavior was observed numerically in~\cite{cagnetta1} for a single-step growth model with discrete activators, which is described by the generic-active-membrane class.

%Summing the structure factor over the modes, the $0$-th excluded, yields the width of an interface having linear size $L$~\cite{krug1997origins},
%\begin{equation}\label{eq:DiscreteToContinuumWidth}
% w^2(L,t) = \frac{1}{L^d}\sum_{\bm{k}\neq 0} S(k,t) \xrightarrow{L,t\to\infty} C \int_{2\pi/L}^\infty k^{d-1}dk\, S(k).
%\end{equation}
%Integrating $k^{d-1}S(k)\sim k^{2y_h-1}$ with respect to $k$, we get the scaling of the saturation ($t\to\infty$) width with the system size
%\begin{equation} \label{eq:Width}
%	w^2_{\infty}(L) \sim L^{-2y_h}.
%\end{equation}
%By comparison with the definition of roughness exponent $\chi$ in kinetic roughening theories, $ w^2_{\infty}(L) \equiv L^{2\chi}$~\cite{krug1997origins}, we identify the scaling dimension $y_h$ with the negative roughness exponent, $-\chi$. According to \autoref{eq:acoustic-scaling}, $\chi=(1-d)/2$, meaning $\chi \leq 0$ $\forall$ $d$, or that the interface is flat in the acoustic regime.

%Similarly, the exponent y φ controls the scaling of the density-density correlations, with structure factor in the acoustic regime S φ (k, t) ' kS φ (kt). It is worth noting the vanishing of the structure factor for k → 0, which is typical of hy-peruniform states [42].

Similar arguments can be applied to fluctuations of the density $\phi$. In this case, the exponent $y_{\phi}$ controls the scaling of the density-density correlation and the scaling of the corresponding time-dependent structure factor in the acoustic regime is given by $S_{\phi}(k,t)\sim k^{2y_\phi-d} {\cal S}_{\phi}(kt)= k{\cal S}_{\phi}(kt)$. The integral of the real-space $\phi$-$\phi$ correlation over a portion of space of linear size $L$ gives a measure of fluctuations of the number of activators, according to the following equation:
\begin{equation}\label{equation-activator-number-fluctuations}
 \delta N^2(L,t) = \int_{[0,L]^d} d^dx \int_{[0,L]^d} d^dy\, \avg{\phi(\bm{x},t)\phi(\bm{y},t)}.
\end{equation}
The right hand side of \ref{equation-activator-number-fluctuations} is proportional to the value of $S_{\phi}(k,t)$ at $k\sim 1/L$ times $L^d$. Then, using the acoustic scaling of $S_{\phi}(k,t)$ and substituting the value of $y_\phi$ from \autoref{eq:acoustic-scaling}, we arrive at $\delta N^2(L,t) \sim L^{d-1}$, in the acoustic limit where $t\sim L$. This is the signature scaling of hyperuniform states~\cite{torquato2003}. A point-pattern or density distribution is termed hyperuniform when its large-scale fluctuations are strongly suppressed---as in the density distribution of a crystal. The early-time hyperuniformity of our activator density results from the activators clumping together in finite-size clusters which are (statistically) uniformly distributed over the system (when starting from homogeneous activators on a flat interface), as discussed in~\cite{cagnetta1} within the context of a one-dimensional lattice model. As with height fluctuations, any non-zero value of the diffusive term, here $\nu_{\phi}$, causes a crossover to diffusive behavior and standard number fluctuations $\delta N^2(L,t) \sim L^d$ for times larger than $t\sim L^2$.

\subsection{Scaling laws in the diffusive regime: the active KPZ model}\label{ssec:diffusivescalinglaws}

%Furthermore, some action terms yield the same algebraic relation between dimensions. Therefore, a few assumptions must be made. The first involves the choice of the dynamic exponent. \mre{I don't understand this sentence:} It should be kept in mind that what is chosen is actually the mean-field dynamic exponent, as the true one might gain perturbative corrections below the upper critical dimension. The typical choice when studying relaxational models such as model A and B is $z\,{=}\,2$~\cite{hoenberghalperin-critical}, also referred to as {\it diffusive} scaling for obvious reasons. 

In the diffusive regime ($z\,{=}\,2$, valid physically for $a_h a_\phi=0$), the assumption of $y_{D_h}\,{=}\,y_{D_\phi}\,{=}\,0$ 
leads to the engineering dimensions summarised in \autoref{tab:engineering-dimensions-KPZ}. In particular, by requiring adimensionality of the terms $\nu_h\int d^dx dt\, \ti{h}\partial_a^2 h$ and $\nu_\phi\int d^dx dt\, \ti{\phi}\partial_a^2 \phi$, we get the following relation between engineering dimensions:
\begin{equation}\label{eq:diffusive-assumption}
 y_{\nu_h} = y_{\nu_\phi} = z-2.
\end{equation}
Thus, as $z\,{=}\,2$, we immediately conclude that the parameters $\nu_h$ and $\nu_\phi$ are marginal in the diffusive regime. The implication at the level of the equations of motion is that the Laplacian terms $\partial_a^2 h$ and $\partial_a^2 \phi$ contribute to the critical behavior of the system.

\begin{table}[h!]
	\def\arraystretch{1.3}
	\centering
	\begin{tabular}{| c || c | c | c | c | c | c | c | c | c |}
		\hline \hline
		$\psi$ & $\ti{h}$ & $h$ & $\ti{\phi}$ & $\phi$ & $\nu_h,\nu_\phi$  & $D_h,D_\phi$ & $a_h,a_\phi$ & $c_h,c_\phi$ & $\lambda,\alpha$ 
		\\  \hline
		$y_\psi$ & $\frac{d+2}{2}$ & $\frac{d-2}{2}$ & $\,\,\,\frac{d}{2}\,\,\,$ & $\,\,\,\frac{d}{2}\,\,\,$ & $ 0 $ & $ 0 $ & $+1$ & $-1$ & $\,\,\frac{2-d}{2}\,\,$
		\\ \hline 
	\end{tabular}
	\caption{Engineering dimensions in the diffusive regime ($z=2$), found under the assumption that $y_{D_h}=y_{D_\phi}=0$, which describes the active KPZ model (for $a_h=a_\phi=0$). The parameter $\kappa$, not included in the table, is irrelevant.}.
	\label{tab:engineering-dimensions-KPZ}
\end{table}

%Having set $z\,{=}\,2$ and $y_{D_h}\,{=}\,y_{D_\phi}\,{=}\,0$ allows us to determine 
%The engineering dimensions of all remaining fields are summarised in \autoref{tab:engineering-dimensions-KPZ}. 
We note that the parameters $a_h$ and $a_\phi$ have a positive dimension. These parameters are therefore relevant and drive the system away from the diffusive critical behavior. However, the leading scaling of the dispersion law~\autoref{eq:dispersion-relation} is diffusive as long as $a_h a_\phi\,{=}\,0$, hence the system is scale invariant in three possible cases: (i) either $a_h\,{=}\,a_\phi=0$, (ii) or $a_\phi\,{=}\,0$ with $a_h$ finite, (iii) or $a_h\,{=}\,0$ with $a_\phi$ finite. These three cases will then correspond to universality classes for the dynamics of active membranes in the diffusive regime. Cases (ii) and (iii) cannot be studied directly here, as one of the coefficients of the linear term would grow without bounds under the RG transformation. We shall discuss them separately in the next two sections, where we will revisit the assumption $y_{D_h}\,{=}\,y_{D_\phi}\,{=}\,0$ made in this section. Here we discuss case (i), which corresponds to the active KPZ model, or the origin of the phase diagram shown in~\autoref{fig:phase-diagram}. The curvature coupling terms can be dropped, as $c_h$ and $c_\phi$ have negative engineering dimension. The nonlinear coupling parameters $\lambda$ and $\alpha$ (the other nonlinear coupling parameter $\kappa$ is always irrelevant in this regime) have engineering dimension $y_\lambda\,{=}\,y_\alpha\,{=}\,(2-d)/2$. The condition $y_\lambda\,{=}\,y_\alpha\,{=}\,0$ identifies the naive upper critical dimension of the model as $d_c\,{=}\,2$.

A minimal active KPZ model can be obtained by the following rescaling of fluctuating fields,
\begin{equation}\label{eq:diff-rescaling}
 h\to\sqrt{\frac{2D_h}{\nu_h}} h ,\quad \phi \to\sqrt{\frac{2D_\phi}{\nu_\phi}} \phi.
\end{equation}
As in the acoustic case, such rescaling, which removes the parameters $D_h$ and $D_\phi$ from the theory, is sensible in a renormalization group calculation because the rescaling factor is marginal and does not change when renormalizing the system. For $d\,{>}\,d_c\,{=}\,2$ the nonlinear couplings  are  irrelevant in the perturbative RG sense and the large-scale properties of the system coincide with those of the following simple pair of equations:
\begin{subequations}\label{eq:act-int-diffusive-meanfield}\begin{align}
 \partial_t h &= \nu_h \partial_a^2 h +\sqrt{\nu_h} \xi_n,\\
 \partial_t \phi &= \nu_\phi \partial_a^2 \phi +\sqrt{\nu_\phi} \partial_a \xi_a.
\end{align}\end{subequations}
In Eqs.~(\ref{eq:act-int-diffusive-meanfield}) height and density fluctuations are decoupled and follow simple diffusion equations, with conservative noise for the density fluctuations. In this case (as in all diffusive regimes) we can use~\autoref{eq:DiscreteToContinuumWidth} to find the steady-state $t\to \infty$ value of the width, $ w^2_{\infty}(L) \equiv L^{2\chi}$~\cite{krug1997origins}, because in this case  the scaling function  ${\cal S}_h$ converges for large times. We can then identify the scaling dimension $y_h$ with the negative roughness exponent, $-\chi$. Height fluctuations therefore display the Edwards-Wilkinson scaling, with roughness exponent $\chi=\frac{2-d}{2}$~\cite{EW}. Similarly, we can use the $t\to\infty$ limit of~\autoref{equation-activator-number-fluctuations} to find that activator density fluctuations obey the scaling $\delta N^2(L) \sim L^{d}$, typical of stochastic point processes. However, at and below $d\,{=}\,2$ the nonlinear terms produce corrections to the mean-field scaling, hence they must be accounted for with a renormalization group procedure going beyond the power counting scheme of this section. The equations to consider for the active KPZ model are thus the following
\begin{subequations}\label{eq:act-int-diffusive-nonlinear-intro}\begin{align}
 \partial_t h &= \frac{\alpha}{2}\phi^2 + \frac{\lambda}{2}(\partial_a h)^2 + \nu_h \partial_a^2 h +\sqrt{\nu_h} \xi_n,\\
 \partial_t \phi &= \lambda\partial_a(\phi\partial_a h)+ \nu_\phi \partial_a^2 \phi +\sqrt{\nu_\phi} \partial_a \xi_a,
\end{align}\end{subequations}
which are studied in \autoref{sec:renormalization-diffusive}. 

%In the following subsection we explore a different possibility: instead of recovering scale invariance by tuning the relevant parameters to zero, we `follow' their divergence under scale transformations with a suitable rescaling of time in the equations of motion. %The implications of a positive engineering dimension are best understood within the scale-transformation picture. A scale transformation of the model can be performed at the level of the action, \autoref{eq:action}, or the equations of motion, \autoref{eq:active-interface-fluctuating}, e.g. by integrating out the high-wavevector (and high-frequency) Fourier components of the fields. In the linearised Gaussian model a scale transformation produces a simple rescaling of the model parameters: a positive engineering dimension implies that the given parameter grows under scale transformation, thus the model is not scale-invariant. There are, at this stage, two possibilities to restore invariance for scale transformation: we can either tune relevant parameters to zero, or `follow' their divergence with a suitable rescaling of time in the equations of motion. The latter option will be explored in \autoref{ssec:acoustic}, whereas we consider the former option in \autoref{ssec:diffusive} and \autoref{sec:renormalization-diffusive}.

\subsection{Scaling laws in the diffusive regime: the curvotactic activators model}\label{ssec:curvotactic}

%\subsection{Active membrane model with curvature coupling: engineering dimensions and renormalization}\label{ssec:active-interface}

The results of \autoref{ssec:diffusivescalinglaws} are based on the assumption $y_{D_h}\,{=}\,y_{D_\phi}\,{=}\,0$. This choice (together with $ z = 2 $), forced us to set $a_h\,{=}\,a_\phi\,{=}\,0$ in order for the system to be critical. When relaxing this assumption, we note that, as anticipated, one of the two noise coefficients $D_h$ and $D_\phi$ should  always be assumed to be marginal, together with one of the couplings $a_h$ or $a_\phi$, otherwise we would end up with a theory where one or both the equations are deterministic (and height or density fluctuations become meaningless). More precisely, we need to either assume that both $D_h$ and $a_\phi$ are marginal, or that both $D_\phi$ and $a_h$ are marginal. In this way, noise is transferred from one equation to the other by the appropriate coupling.  %We note that having a single stochastic term is sufficient to introduce noise in the other equation as well through the coupling terms.
%Let us first consider the alternative assumption that only $y_{D_h}=0$ However, because of the non-dissipative couplings $a_h$ and $a_\phi$, dropping only one of the noise terms does not spoil the fields' stochasticity and the whole rationale behind the action formalism. For instance, if $D_h\,{=}\,0$,  the height field $h$ is indirectly affected by the density noise $D_\phi$ through the $a_h\phi$ term in the height equation. Therefore, the assumption $y_{D_h}\,{=}\,0$ could be replaced with $y_{a_h}\,{=}\,0$. This possibility is explored in \autoref{ssec:active-interface}. Alternatively, one could drop the density noise $D_\phi$ and still have a stochastic density field through the $a_\phi\partial_a^2 h$ term and the stochastic term in the height equation. This other case, highlighted when the assumption $y_{D_\phi}\,{=}\,0$ is replaced with $y_{a_\phi}\,{=}\,0$, is examined in \autoref{ssec:passive-sliders}. Unfortunately, none of these new regimes displays new perturbative fixed points below the upper critical dimension---$d_c\,{=}\,4$ in the first case, $d_c\,{=}\,2$ in the second. Nevertheless, our analysis shows that, when the linear non-dissipative couplings are silenced by tuning the corresponding parameter to zero, the dissipative curvature-coupling becomes relevant for the scaling properties of the model.

We first consider the case in which $y_{D_\phi}\,{=}\,y_{a_h}\,{=}\,0$. This allows us to describe the $a_\phi\,{=}\,0$ line of the phase diagram---corresponding to the orange line in \autoref{fig:phase-diagram}, or to the curvotactic activators model. Here, fluctuations of height and density show another kind of critical behavior. The engineering dimensions of this regime are summarised in \autoref{tab:engineering-dimensions-activeh}.

\begin{table}[ht!]
	\def\arraystretch{1.3}
	\centering
	\begin{tabular}{| c || c | c | c | c | c | c | c | c | c |}
		\hline \hline
		$\,\psi\,$ & $\ti{h}$ & $h$ & $\,\,\,\ti{\phi}\,\,\,$ & $\,\,\,\phi\,\,\,$ & $\nu_h,\nu_\phi$  & $\,a_h,D_\phi,c_\phi\,$ & $a_\phi$ & $D_h,c_h$ & $\,\lambda,\kappa\,$ 
		\\  \hline
		$y_\psi$ & $\frac{d+4}{2}$ & $\frac{d-4}{2}$ & $\frac{d}{2}$ & $\frac{d}{2}$ & $ 0 $ & $ 0 $ & $+2$ & $-2$ & $\frac{4-d}{2}$
		\\ \hline 
	\end{tabular}
	\caption{Engineering dimensions in the diffusive regime ($z=2$), found under the assumption that $y_{D_\phi}=0$, which describes the case of the curvotactic activators model ($a_\phi=0$ line in Fig.~\ref{fig:phase-diagram}).}
	\label{tab:engineering-dimensions-activeh}
\end{table}

The salient features shown in the table are that (i) the only parameter which is always relevant is $a_\phi$, with the engineering dimension of a mass, (ii) $D_h$ and $c_h$, i.e. noise and curvature-coupling in the height equation, are irrelevant, (iii) the dimensions of the nonlinear couplings $\lambda$ and $\kappa$ vanish at $d=4$. Thus, the naive upper critical dimension is $d_c\,{=}\,4$ for this case. The minimal system of equations to describe the curvotactic activators model is obtained by the following rescaling of fields,
\begin{equation}\label{eq:rescaling-activeh}
h\to \sqrt{\frac{2D_\phi}{\nu_\phi}}\frac{a_h}{\nu_h} h , \quad \phi \to \sqrt{\frac{2D_\phi}{\nu_\phi}} \phi.
\end{equation}
Above $d\,{=}\,4$, where the mean-field scaling of~\autoref{tab:engineering-dimensions-activeh} should hold exactly, the nonlinear terms can be dropped and the large-scale properties of the model (at the scale of \autoref{eq:rescaling-activeh}) are those of the following pair of equations,
\begin{subequations}\label{eq:a-i-e-activeh}
\begin{align}\label{eq:a-i-activeh-int}
\partial_t h &=  \nu_h\phi + \nu_h \partial_a^2 h, \\ 
\label{eq:a-i-activeh-dens}
\partial_t \phi &= - c_\phi' \partial_a^4 h + \nu_\phi \partial_a^2 \phi+ \partial_a\left(\sqrt{\nu_\phi}\xi_a\right),
\end{align}
\end{subequations}
where $c_\phi'\,{=}\, c_\phi a_h/\nu_h$.

The equations above entail, essentially, the same ingredients of the model discussed in~\cite{gov2006}. According to our analysis, the interaction between the activator positions and the interface curvature becomes marginal, therefore important for the scaling of the system, only in the present case, when the kinematic coupling with the slope, $a_\phi$, is made to vanish. In addition, in this regime, the stochastic fluctuations of the interface are masked by the active fluctuations coming from the activator distribution. Thus, the roughness exponent is $\chi=\frac{4-d}{2}$, which is different from that of the Edwards-Wilkinson model, and this is due to the noise transferred from the $\phi$ equation. Density fluctuations instead obey the standard central limit theorem scaling, $\delta N^2(L) \sim L^{d}$. Below $d\,{=}\,4$ the nonlinear couplings $\lambda$ and $\kappa$ become relevant. A full renormalization group analysis is required here, and we discuss it in \autoref{ssec:curvotactic-renormalization}. 
%the equations to study are
%\begin{subequations}\label{eq:a-i-e-activeh}
%\begin{align}\label{eq:a-i-activeh-int}
%\partial_t h &=  \nu_h\phi + \nu_h \partial_a^2 h, \\ 
%\label{eq:a-i-activeh-dens}
%\partial_t \phi &= - c_\phi' \partial_a^4 h + \nu_\phi \partial_a^2 \phi+ \partial_a\left(\sqrt{\nu_\phi}\xi_a\right).
%\end{align}
%\end{subequations}

\subsection{Scaling laws in the diffusive regime: the passive sliders model}\label{ssec:passivesliders}

The scale-invariant properties of the model on the $a_h\,{=}\,0$ line, where also $z=2$, are highlighted by assuming $y_{D_h}\,{=}\,y_{a_\phi}\,{=}\,0$. This is the passive sliders model, corresponding to the blue line in \autoref{fig:phase-diagram}. The engineering dimensions corresponding to this regime are summarised in \autoref{tab:engineering-dimensions-passivep}.
\begin{table}[h!]
	\def\arraystretch{1.3}
	\centering
	\begin{tabular}{| c || c | c | c | c | c | c | c | c | c |}
		\hline \hline
		$\,\psi\,$ & $\ti{h}$ & $h$ & $\,\,\,\ti{\phi}\,\,\,$ & $\,\,\,\phi\,\,\,$ & $\nu_h,\nu_\phi$  & $\,a_\phi,D_h,c_h\,$ & $a_h$ & $D_\phi,c_\phi$ & $\,\lambda\,$ 
		\\  \hline
		$y_\psi$ & $\frac{d+2}{2}$ & $\frac{d-2}{2}$ & $\frac{d+2}{2}$ & $\frac{d-2}{2}$ & $ 0 $ & $ 0 $ & $+2$ & $-2$ & $\frac{2-d}{2}$
		\\ \hline 
	\end{tabular}
	\caption{Engineering dimensions in the diffusive regime ($z=2$), found under the assumption that $y_{D_h}=0$, which describes the passive slider model ($a_h=0$ line in Fig.~\ref{fig:phase-diagram}).
}
	\label{tab:engineering-dimensions-passivep}
\end{table}

In the passive sliders model %described by \autoref{tab:engineering-dimensions-activeh} 
$a_h$ is the only parameter which is always relevant. As $D_\phi$ and $c_\phi$ have negative engineering dimensions, the density noise and the curvature coupling in the density equation  are irrelevant. The dimension of the nonlinear coupling $\lambda$, as in the model of \autoref{sec:powercounting}, vanishes at $d\,{=}\,2$. A minimal version of the model can be obtained by considering the following field rescaling
\begin{equation}\label{eq:rescaling-pasivep}
h\to \sqrt{\frac{2D_h}{\nu_h}} h , \quad \phi \to \sqrt{\frac{2D_h}{\nu_h}} \frac{a_\phi}{\nu_\phi}\phi,
\end{equation}
which causes the coefficients of the Laplacian term $\partial_a^2 \phi $ and that of the slope-coupling term $\partial_a^2 h$ in the density equation to coincide. Above $d\,{=}\,2$, the mean-field scaling of~\autoref{tab:engineering-dimensions-activeh} holds and the large-scale properties of the model are described by the following pair of equations,
\begin{subequations}\label{eq:a-i-e-passivep}
\begin{align}\label{eq:a-i-passivep-int}
\partial_t h &=  -c'_h\partial_a^2 \phi + \nu_h \partial_a^2 h + \sqrt{\nu_h}\xi_n, \\ 
\label{eq:a-i-passivep-dens}
\partial_t \phi &= \nu_\phi \partial_a^2 h + \nu_\phi \partial_a^2 \phi,
\end{align}
\end{subequations}
where  $c_h'\,{=}\, c_h a_\phi/\nu_\phi$.

Interestingly, tuning the coefficient of the active force term to zero renders the curvature coupling $c_h$ a marginal term with respect to the large-scale properties of the active interface model. This term models the action of membrane proteins which, rather than causing displacement of the interface, impose a given local curvature on the region where they sit. Another interesting feature of this passive sliders regime is that the stochastic fluctuations of the activators within the membrane are masked by those caused by the stochastic motion of the membrane. Thus, the interfacial fluctuations, which are described by an EW-like roughness exponent, $\chi=\frac{2-d}{2}$, create anomalous fluctuations in the density, described by a scaling $\delta N^2(L) \sim L^{d+2}$, larger than expected from the central limit theorem. We conclude that interfacial fluctuations induce {\it giant number fluctuations} in the density of activators~\cite{chakraborty2020}. Below $d\,{=}\,2$ the nonlinear terms should be taken into account with a renormalization group procedure which we discuss in \autoref{ssec:passivesliders-renormalization}. %However, the model turns out to be non-renormalizable, as there are an infinite number of diverging perturbative corrections. Recalling \autoref{eq:primitive-degree-divergence-diffusive} for the primitive degree of divergence of perturbative corrections, using the engineering dimensions of \autoref{tab:engineering-dimensions-passivep} and $d\,{=}\,2$,
%\begin{equation}\label{prim-deg-div-passivep}
% \delta(\mathcal{I}) = 4 - 2n_{\ti{h}} -2 n_{\ti{\phi}}.
%\end{equation}
%The diagrams containing diverging contributions are again those for which the primitive degree of divergence exceeds the number of derivatives appearing in the corresponding term of the action. The vertex functions which satisfy this condition at $d\,{=}\,4$ are found to be
%\begin{equation}\label{eq:vertex-functions-pasivep}\begin{aligned}
% \Gamma^{\ti{h}h\phi^j},\quad
% \Gamma^{\ti{h}hh\phi^k} ,\quad
% \Gamma^{\ti{h}h\phi^l},\quad
% \Gamma^{\ti{\phi}\phi\phi^m},\quad \Gamma^{\ti{\phi}h\phi^n},
%\end{aligned}\end{equation}
%for any integer value of $j,k,l,m$ and $n$. There are then $5$ infinite families of terms that would need to be included in the theory in order to perform renormalization around $d\,{=}\,2$. This is caused by the engineering dimension of the field $\phi$ which, by vanishing at $d\,{=}\,2$, implies a breakdown of the small $\phi$ approximation used in deriving the model. Similar models (albeit devoid of the curvature term proportional to $c_h$) have been considered before, with no clear scaling picture emerging~\cite{das2000,drossel2002}. \autoref{prim-deg-div-passivep} suggests that this could be due to the non-renormalizability of the model: non-renormalizability can indeed be associated with non-universal behavior~\cite{antonov2017}.

\section{renormalization of the active KPZ model}\label{sec:renormalization-diffusive}

Let us now discuss in detail the scaling and the critical properties of the active KPZ model at the $a_h\,{=}\,a_\phi\,{=}\,0$ point of the parameter space. The equations to consider in this case read (\autoref{eq:act-int-diffusive-nonlinear-intro})
\begin{subequations}\label{eq:act-int-diffusive-nonlinear}\begin{align}
		\partial_t h &= \frac{\nu\alpha}{2}\phi^2 + \frac{\nu\lambda}{2}(\partial_a h)^2 + \nu \partial_a^2 h +\sqrt{\nu} \xi_n;\\
		\partial_t \phi &= \nu\lambda\partial_a(\phi\partial_a h)+ r\nu \partial_a^2 \phi +\sqrt{r\nu} \partial_a \xi_a,
	\end{align}
\end{subequations}
where we have set $\nu_h = \nu$ and $\nu_\phi = r\nu$, with $r$ measuring the ratio between the ``viscosities'' $\nu_h$ and $\nu_\phi$. We have also conveniently rescaled the coefficients of the nonlinear terms, $\alpha$ and $\lambda$, with $\nu$. Naive power counting tells us that the nonlinear terms influence the critical properties of the model for $ d \leq 2 $. In this section we study the corresponding shift of the critical exponents from the mean-field values of~\autoref{tab:engineering-dimensions-KPZ}, with the RG method. The main idea is to perform a systematic renormalization of the UV divergences at the upper critical dimension $d_c$ which, in turn, allows us to study the large scale, i.e. \emph{infrared} (IR) limit, in the vicinity of $d_{c}$ due to the relation between UV and IR limit at $ d\,{=}\,d_{c}$~\cite{amitFieldTheory,tauberCriticalDynamics,Vasilev04}.

\subsection{Primitive degree of divergence of vertex functions, and renormalizability of the model}

In order to show that the model is indeed renormalizable at the upper critical dimension, let us examine perturbative corrections to a generic vertex functions. An $n$-point vertex function $\Gamma^{\psi_1\dots\psi_n}$ is obtained by differentiating the effective action \eqref{eq:effective-action} of \autoref{eq:vertex-gen-fun} $n$ times with respect to the fields. Therefore, its scaling dimension is given by the negative sum of the scaling dimensions of the fields $\psi_i$, plus a factor $d+z\,{=}\,d+2$ from the $\delta$-function which imposes conservation of momentum. Perturbative corrections to $\Gamma^{\psi_1\dots\psi_n}$ generically include some power of the nonlinear couplings $\alpha$ and $\lambda$ multiplied by a momentum integral, which we denote by $\mathcal{I}^{\psi_1\dots\psi_n}$. Since perturbative corrections must have the same dimension of the vertex function itself, the engineering dimension of $\mathcal{I}^{\psi_1\dots\psi_n}$ must be
\begin{equation}\label{eq:primitive-degree-divergence-diffusive}
 \delta(\mathcal{I}^{\psi_1\dots\psi_n}) = d+2 -\sum_{i=1}^n y_{\psi_i} -n_\lambda y_\lambda - n_\alpha y_\alpha,
\end{equation}
where $n_\lambda$ and $n_\alpha$ denotes the number of $\lambda$ and $\alpha$ factors in the perturbative correction---i.e., the order of the perturbation.

The left-hand side of \autoref{eq:primitive-degree-divergence-diffusive} is called the {\it primitive degree of divergence} of the graph associated with $\mathcal{I}^{\psi_1\dots\psi_n}$. By definition, this quantity determines the power of the superficial UV divergence of perturbative corrections to a given vertex function. For the theory to be renormalizable at an integer dimension $d$, the number of divergent vertex functions must be finite and independent of the order of perturbation theory. The last condition, according to \autoref{eq:primitive-degree-divergence-diffusive}, requires $y_\alpha,y_\lambda\,{=}\,0$ which means that the theory is renormalizable at $ d \,{=}\, d_{c}\,{=}\,2 $, where the relation between UV and IR limits exists. Results can then be analytically continued to the  vicinity of $d_c$. Setting $d\,{=}\,2$ in the engineering dimensions of the fields (taken from \autoref{tab:engineering-dimensions-KPZ}) gives
\begin{equation}\label{eq:prim-deg-div-ucd-KPZ}
 \delta(\mathcal{I}^{\psi_1\dots\psi_n}) = 4 -2 n_{\ti{h}} - n_\phi- n_{\ti{\phi}}.
\end{equation}
There are infinite vertex functions with a positive $\delta$. However, one must take into account that there are gradients entering $\mathcal{I}^{\psi_1\dots\psi_n}$: as these do not cause UV divergences, such gradients need to be factored out when assessing renormalizability\cite{Vasilev04,Skultety2021}.
%one power of momentum comes from the coupling between the gradient and $h$, and one more from the coupling between the other gradient and the first, which is required by rotational invariance.
In general, each $h$ field appearing in the theory comes coupled with a gradient, because of translational invariance along the $h$ direction; $\ti{\phi}$'s are also coupled to gradients because of the density conservation, and each solitary gradient must contract with another gradient to preserve rotational invariance. As a consequence, for instance $\Gamma^{\ti{h}h}$ has $\delta\,{=}\,2$, but also two gradients, so that the degree associated with the factor multiplying the gradients is $0$. The same is true of $\Gamma^{\ti{\phi}\ti{\phi}}$. The important vertex functions to consider are therefore those for which $\delta \geq n_{\nabla}$, with $n_\nabla$ denoting the number of spatial gradients the corresponding term of the action, or twice the number of time derivatives (as $z=2$). The only
vertex functions satisfying these conditions are
  %For instance, $\Gamma^{\ti{h}h}$ has $\delta\,{=}\,2$, but one power of momentum comes from the coupling between the gradient and $h$, and one more from the coupling between the other gradient and the first, which is required by rotational invariance. Analogously, the degree of divergence of $\Gamma^{\ti{\phi}\ti{\phi}}$, which is $2$, comes from the two gradients coupled with each $\phi$ field, and the same applies to $\Gamma^{\ti{\phi}\phi}$, $\Gamma^{\ti{h}h}$,  $\Gamma^{\ti{h}hh}$ and $\Gamma^{\ti{\phi}\phi h}$. In general, rather than listing vertex functions with positive $\delta$, one should list the vertex functions with $\delta \geq n_{\nabla}$, with $n_\nabla$ denoting the number of gradients in the corresponding term of the action (or twice the number of time derivatives, because $z=2$). The only vertex functions satisfying this condition are
\begin{equation}\label{eq:divergent-vertices-KPZ}
 \Gamma^{\ti{h}\ti{h}},\quad \Gamma^{\ti{h}h},\quad \Gamma^{\ti{\phi}\ti{\phi}},\quad \Gamma^{\ti{\phi}\phi},\quad \Gamma^{\ti{h}hh},\quad \Gamma^{\ti{\phi}\phi h},\quad \Gamma^{\ti{h}\phi \phi},
\end{equation}
which are the ones required to derive the RG equations and flow for the active KPZ model.
All the vertex functions  not listed in~\autoref{eq:divergent-vertices-KPZ}, are not affected by UV-divergent perturbative corrections. Therefore, the associated parameters are irrelevant in the RG sense and do not affect the scaling of the model.

The identification of the exact form of divergent structures proceeds in a similar fashion as above. We know, for instance, that every divergent contribution to $\Gamma^{\ti{h}h}$ must contain an extra $\partial^2$ operator, hence the diffusive term will be renormalized while the frequency term receives no corrections (cf.~\autoref{eq:KPZ-Ghth-oneloop}). Moreover, as the tilt symmetry~\autoref{eq:tilt-transformation} enforces the frequency term of $\Gamma^{\ti{h}h}$ to gain identical corrections as $\Gamma^{\ti{h}hh}$, the latter is unaffected by the renormalization process. Similar ideas apply to the density part of the model: in conclusion, the only terms that require renormalization are
\begin{equation}
 \ti{h}\ti{h},\quad \ti{h}\partial^2 h,\quad \ti{\phi}\partial^2\ti{\phi},\quad \ti{\phi}\partial^2 \phi,\quad \ti{h}\phi^2,
\end{equation}
at all orders of the loop expansion. Let us remark that this does not imply that we may simply omit the KPZ nonlinearity and the analogous one in the density equation: they are both marginal in the RG sense and the coupling parameter $\lambda$ is still renormalized due to the renormalization of the fields.

\subsection{One-loop renormalization of perturbative corrections}

One-loop corrections to the vertex functions listed in~\autoref{eq:divergent-vertices-KPZ} are computed in~Appendix B %\autoref{app:perturbationKPZ} 
(Eqs.~\eqref{eq:KPZ-Ghth-oneloop}-\eqref{eq:KPZ-Ghtpp-oneloop}). Specifically, dimensional regularisation was used to evaluate the integrals for any real value of the spatial dimension $d$, so that UV divergences at $ d_{c}\,{=}\,2 $ appear as $ 1/(2-d) $ poles. The goal of this section is to absorb these divergent contributions by defining a set of renormalized parameters and fields. To this end, it is convenient to denote the coefficients of the bare theory with a subscript $_0$, in order to distinguish them from the renormalized coefficients (with no subscript). We define renormalized quantities as
\begin{equation}\label{eq:dimensionless-coupling-KPZ}
 \psi_0 = Z_\psi \psi, \quad \lambda_0 =\frac{\mu^{\frac{2-d}{2}}}{\sqrt{S_d}} Z_\lambda \lambda ,\quad \alpha_0 = \frac{\mu^{\frac{2-d}{2}}}{\sqrt{S_d}} Z_\alpha \alpha,
\end{equation}
where $\psi$ denotes a generic field/parameter other than a nonlinear coupling constant. The arbitrary momentum scale $\mu$ has been introduced in order to absorb the engineering dimensions of the bare nonlinear couplings $\lambda_0$ and $\alpha_0$, so that the renormalized couplings are dimensionless. The renormalization constants $ Z $ are chosen so as to incorporate the UV-divergent corrections in ~\hyperref[eq:KPZ-Ghth-oneloop]{Eqs.~(\ref*{eq:KPZ-Ghth-oneloop}-\ref*{eq:KPZ-Ghtpp-oneloop})}. The effective action \eqref{eq:effective-action} then becomes
\begin{widetext}\begin{equation}\label{eq:renormalized-action-KPZ}
%%%%%%%%%%%%%%%%%%%%%%%%%%%%%%%%%%%%%%%%%%%%%%%%%
\begin{aligned}
\Gamma[\ti{h},h,\ti{\phi},\phi] &=\int d^d\bm{x}dt\, \left[\ti{h}_{0} \left(\partial_t - \nu_0 \nabla^2 \right)h_{0} + \tilde{\phi}_{0}\left(\partial_t -r_0\nu_0 \nabla^2 \right)\phi_{0} - \frac{\nu_0}{2}\ti{h}_{0}^2 + \frac{r_0\nu_0}{2}\ti{\phi}_{0}\nabla^2 \ti{\phi}_{0} \right. \\ 
%%%%%%%%%%%%%%%%%%%%%%%%%%%%%%%%%%%%%%%%%%%%%%%%%
&\left. -\frac{\nu_0\alpha_0}{2}\ti{h}_{0}\phi_{0}^2-\frac{\nu_0\lambda_0}{2}\ti{h}_{0}\left(\nabla h_{0}\right)^2 + \nu_0\lambda_0 \left(\bm{\nabla}\ti{\phi}_{0}\right)\cdot \left(\phi_{0}\bm{\nabla}h_{0}\right) \right] + \text{one-loop corrections} 
\\
%%%%%%%%%%%%%%%%%%%%%%%%%%%%%%%%%%%%%%%%%%%%%%%%%
&\equiv \int d^d\bm{x}dt\, \left[ \ti{h}\left(\partial_t - \nu \nabla^2 \right)h + \tilde{\phi}\left(\partial_t - r\nu \nabla^2 \right)\phi - \frac{\nu}{2}\ti{h}^2 + \frac{r\nu}{2}\ti{\phi}\nabla^2 \ti{\phi} \right. \\ 
%%%%%%%%%%%%%%%%%%%%%%%%%%%%%%%%%%%%%%%%%%%%%%%%%
&\left. - \mu^{\frac{2-d}{2}}\nu \left(\frac{\alpha}{2}\ti{h}\phi^2- \frac{\lambda}{2}\ti{h}\left(\nabla h\right)^2 + \lambda \left(\bm{\nabla}\ti{\phi}\right)\cdot \left(\phi\bm{\nabla}h\right) \right)\right] + \text{UV finite corrections}
\end{aligned}
%%%%%%%%%%%%%%%%%%%%%%%%%%%%%%%%%%%%%%%%%%%%%%%%%
\end{equation}\end{widetext}
The exact form of the renormalization constants is obtained by adopting the \emph{minimal subtraction scheme} (MS)---i.e., by demanding the effective action to be UV-finite in the limit $ \epsilon = 2 - d \rightarrow 0 $,  which gives
\begin{flalign}\label{eq:r-c-KPZ-Znu} &\quad Z_\nu = 1 - \frac{\alpha\lambda}{8r\epsilon};& \end{flalign} \begin{flalign}\label{eq:r-c-KPZ-Zr} &\quad Z_r = 1- \frac{(1-r)\left[2\lambda^2+(r-1)\alpha\lambda\right]}{8r(1+r)^2\epsilon};& \end{flalign} \begin{flalign}\label{eq:r-c-KPZ-Zhth} &\quad Z_{\ti{h}} = 1-\frac{r\lambda^2+\alpha^2-\alpha\lambda}{16r\epsilon},\quad Z_h = 1+\frac{r\lambda^2+\alpha^2-\alpha\lambda}{16r\epsilon}; &\end{flalign} \begin{flalign} \label{eq:r-c-KPZ-Zptp} &\quad Z_{\ti{\phi}} = 1 -\frac{\lambda^2 -\alpha\lambda}{4(1+r)^2\epsilon}, \quad Z_\phi = 1 +\frac{\lambda^2 -\alpha\lambda}{4(1+r)^2\epsilon}; &\end{flalign} \begin{flalign}\label{eq:r-c-KPZ-Zlambda} &\quad Z_\lambda = 1-\frac{r\lambda^2+\alpha^2-3\alpha\lambda}{16r\epsilon};&\end{flalign} \begin{flalign}\label{eq:r-c-KPZ-Zalpha} &\quad \begin{aligned}Z_\alpha = 1 +\frac{(1+r)^2 \alpha^2 + \left(9 +10r -7r^2 \right)\alpha\lambda}{16r(1+r)^2\epsilon} \\  + \frac{\left(r^3 + 10r^2 -7r-8\right)\lambda^2}{16r(1+r)^2\epsilon}.\end{aligned}&\end{flalign}

\subsection{RG flow and universality}

We can now write down the RG equations, which probe the universal scaling of the system in the IR limit.
%Once the theory has been renormalized the system becomes scale dependent. The RG method then allows us to probe the universal scaling in the IR limit. 
To do so, we note that the following relation holds between the bare and renormalized two-point correlation functions for two generic fields $\psi_{i}$ and $\psi_{j}$,
\begin{equation}\label{eq:functional-RG-equation}
C_0^{\psi_{i}\psi_{j}} (\Bbbk, \nu_0, g_{0},m_{0}) = Z_{\psi_{i}} Z_{\psi_{j}} C^{\psi_{i}\psi_{j}} (\Bbbk, \nu, g,m,\mu),
\end{equation}
where $g\,{=}\,\{\lambda,\alpha,r\} $. As the bare theory is independent of the arbitrary momentum scale $\mu$, the logarithmic derivative with respect to $\mu$ of \autoref{eq:functional-RG-equation} yields
\begin{align} \label{eq:RG-equation}
\bigg( \mu \frac{\partial}{\partial \mu} + \beta_{g} \partial_{g} - \gamma_{\nu} & \nu \frac{\partial}{\partial \nu} + \gamma_{\psi_{i}} + \gamma_{\psi_{j}} \bigg) \nonumber \\
&\times C^{\psi_{i}\psi_{j}} (k,\omega, g,m,\mu) = 0,
\end{align}
where we have introduced beta functions and anomalous dimensions as
\begin{align}
%%%%%%%%%%%%%%%%%%%%%%%%%%%%%%%%%%%%%%%%%%%%%%%%%
\beta_{g} = - g \left( \frac{\epsilon}{2} + \gamma_{g} \right), \quad \gamma_{\psi} = \mu \frac{\partial}{\partial \mu} \bigg|_{\text{bare}} \ln Z_{\psi}.
%%%%%%%%%%%%%%%%%%%%%%%%%%%%%%%%%%%%%%%%%%%%%%%%%
\end{align}
Here $\psi$ denotes a generic field/parameter and derivatives are taken with the bare parameters $\psi_0$ fixed. %Note the absence of the $ m\partial_{m} $ operation in \eqref{eq:RG-equation} due to the non-renormalization of the IR cut-off parameter $m$.

Considering the $Z$ factors in \autoref{eq:r-c-KPZ-Zr}, \autoref{eq:r-c-KPZ-Zlambda} and \autoref{eq:r-c-KPZ-Zalpha}, and replacing $\alpha$ and $\lambda$ with the bare counterparts $\mu^{\frac{d-2}{2}}\lambda_0$ and $\mu^{\frac{d-2}{2}}\alpha_0$ (additional $Z$ factors are set to $1$ in order to keep the perturbative expansion first order in the number of loops), the following beta functions are obtained:
\begin{subequations}\label{eq:beta-functions-KPZ}\begin{align}
  \label{eq:b-f-KPZ-r} \beta_r &=\frac{(r-1)\left[2\lambda^2+(r-1)\alpha\lambda\right]}{8(1+r)^2};\\
  \label{eq:b-f-KPZ-lambda} \beta_\lambda &= -\frac{\lambda}{2}\left(\epsilon+\frac{r\lambda^2+\alpha^2-3\alpha\lambda}{8r} \right);\\
  \label{eq:b-f-KPZ-alpha} \beta_\alpha &=  \frac{\alpha}{2}\left(\epsilon-\frac{(1+r)^2 \alpha^2 +\left(9 +10r -7r^2 \right)\alpha\lambda}{8r(1+r)^2}\right.\nonumber\\ &+\left. \frac{\left(r^3 + 10r^2 -7r-8\right)\lambda^2}{8r(1+r)^2}\right).
\end{align}\end{subequations}
Because of the definition of beta functions, their zeroes are the possible fixed points of a scale transformation parameterised by the momentum scale $\mu$---i.e., of the RG flow. The fixed point stability is determined from the eigenvalues of the matrix $\partial_i \beta_j \equiv \partial_{g_i} \beta_{g_j}$. In the IR limit $k/\mu, \omega/\nu\mu^{2},m/\mu\to 0$, so that a positive eigenvalue signifies stability.

The first, trivial, zero of \autoref{eq:beta-functions-KPZ} is $\lambda\,{=}\,\alpha\,{=}\,0$. This corresponds to the Gaussian fixed point, for which $r$ can assume any value. This fixed point is stable for $\epsilon\,{<}\,0$, i.e. $d\,{>}\,d_c$, as expected from power counting. Setting $\lambda\,{=}\,0$ yields two additional fixed points with $\alpha\,{=}\,\pm 8r\epsilon$. Such fixed points only exist for $\epsilon\,{<}\,0$---i.e., $d\,{>}\,d_c$---and are saddle points. With $\lambda\,{\neq}\, 0$, $\beta_r$ vanishes also at $r\,{=}\,1$ or $r\,{=}\,(\alpha-2\lambda)/\alpha$. Setting $r\,{=}\,1$ yields the KPZ roughening transition fixed point discussed in~\cite{tauberCriticalDynamics}, with $\lambda^2\,{=}\,-8\epsilon$ and $\alpha=0$, which exists only for $\epsilon\,{<}\,0$ and is unstable. Setting $r\,{=}\,(\alpha-2\lambda)/\alpha$ does not yield any other fixed points in the physical sector $r\,{>}\,0$. The fixed points of the RG flow are summarised in~\autoref{tab:f-p-diffusive}, together with the corresponding eigenvalues of the linearised RG flow.
\begin{table}[ht!]
	\def\arraystretch{1.3}
	\centering
	\begin{tabular}{| c || c | c | c || c | c | c |}
		\hline \hline
		$\,$ & $\quad r\quad$ & $\lambda$ & $\alpha$ & $e_1$ & $e_2$ & $e_3$ \\  \hline
		$P_0 \,(GAUSS)$ & $\text{any}$ & $0$ & $0$ & $0$ & $-\epsilon/2$ & $-\epsilon/2$ \\ \hline
		$P_1 \,(KPZ)$ & $1$ & $\pm\sqrt{-8\epsilon}$ & $0$ & $-\epsilon/2$ & $ \epsilon $ & $0$ \\ \hline
		$P_2 \,(NEW)$ & $\text{any}$ & $0$ & $\pm\sqrt{8r\epsilon}$ & $0$ & $ -\epsilon $ & $\epsilon$  \\ \hline
	\end{tabular}
	\caption{Fixed points of the RG flow at the diffusive scale. The stability of the fixed point is determined by the sign of the eigenvalues of $\partial_i \beta_j$ at the fixed point, with $i,j\,{=}\,r,\lambda,\alpha$. Because the relevant limit is the infrared limit (small-momenta/large-scales), a positive eigenvalue is sign of stability of the corresponding fixed point.}.
	\label{tab:f-p-diffusive}
\end{table}

\begin{figure*}[t!]
\begin{center}
  \begin{tabular}{cc}
       \includegraphics[width=2.0\columnwidth]{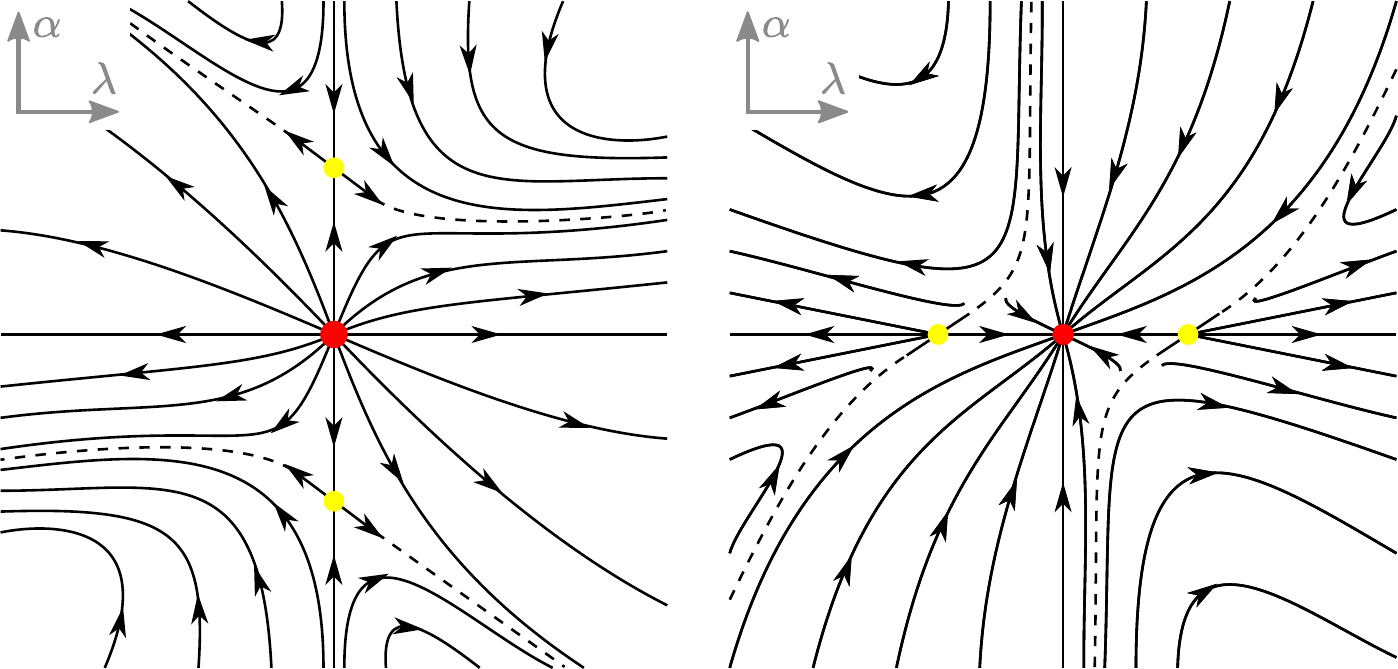}\\
  \end{tabular}
\caption{Phase diagram in the $\alpha$-$\lambda$ plane of the model described by \autoref{eq:act-int-diffusive-nonlinear}. For $d\,{<}\,2$ (a), there is a perturbative fixed point on the $\lambda\,{=}\,0$ line, marked by the yellow dot. For $d\,{>}\,2$ (b) the flow converges onto the Gaussian fixed point along the $\alpha$ direction. The yellow dot on the $\alpha\,{=}\,0$ line marks the IR-unstable fixed point which marks the famous roughening transition of the KPZ equation.}
\label{fig:phase-diagram-KPZ}
\end{center}
\end{figure*}

Let us now discuss the three fixed points in more detail, in the $r-\lambda-\alpha$ space. Recall that, since $a_h\,{=}\,a_\phi\,{=}\,0$, we are studying the origin of the reduced phase diagram shown in \autoref{fig:phase-diagram}. We will express directions in the $r-\lambda-\alpha$ space in terms of the canonical basis $(\bm{e}_r,\bm{e}_\lambda,\bm{e}_\alpha)$. The first fixed point we discuss is the Gaussian fixed point $P_0$: this is the red point in \autoref{fig:phase-diagram-KPZ}a,b. At $P_0$, the matrix $\partial_i \beta_j$ has two eigenvalues: one is zero, and the corresponding eigenvector is parallel to $\bm{e}_r$. The other, degenerate, is proportional to $-\epsilon$ and its eigenvectors span the $\lambda-\alpha$ plane. As a result, the Gaussian fixed point is marginal for all $d$ in the $r$ direction, while, in the $\lambda-\alpha$ plane, it is repulsive for $d\,{<}\,2$ and attractive for $d\,{>}\,2$ (see \autoref{fig:phase-diagram-KPZ}).

The second fixed point, $P_1$, is that corresponding to the roughening transition of the KPZ equation in $d>d_c=2$~\cite{tauberCriticalDynamics} and is marked by a yellow dot in \autoref{fig:phase-diagram-KPZ}b. $P_1$ is a saddle-point: of the three eigenvalues of $\partial_i \beta_j$, one is positive (with eigenvector $\propto\bm{e}_r$), one is negative (with eigenvector $\propto\bm{e}_\lambda$) and one is zero (with eigenvector $\propto 3\bm{e}_\lambda+2\bm{e}_\alpha$). Therefore, the roughening transition fixed point is attractive in the $r$ direction and repulsive in the $\lambda$ direction. For $\alpha\,{=}\,0$, the interface will be rough in steady state for $|\lambda|> |\lambda_c|\,{=}
\,\sqrt{8(d-2)}$ and smooth otherwise. Because of the marginal direction $3\bm{e}_\lambda+2\bm{e}_\alpha$, having a small $\alpha$ shifts the roughening threshold $\lambda_c$ by $3\alpha /2$.

The third fixed point $P_2$ is the most important one in our analysis, as it represents a new kind of universal behavior for active interfaces. $P_2$ corresponds to the two yellow points in \autoref{fig:phase-diagram-KPZ}a,  with identical critical exponents, and it only exists below $d\,{=}\,2$. It has a marginal direction ($\bm{e}_r$, with zero eigenvalue), an attractive direction ($\bm{e}_\alpha$, with eigenvalue $\epsilon$) and a repulsive one ($3\bm{e}_\alpha-4\bm{e}_\lambda$, with eigenvalue $-\epsilon$). The corresponding flow is depicted in \autoref{fig:phase-diagram-KPZ}a. For nonzero $\lambda$, the flow drives the model towards larger $\lambda$'s, possibly to a non-perturbative fixed point analogous to the one which describes the KPZ scaling in one dimension~\cite{Canet2010,Canet2011}. However, if $\lambda$ is tuned to zero, the flow converges onto $\alpha^2= 8r(2-d)$, with $r$ arbitrary. Because $\lambda$ is related to the average speed of the membrane, this new fixed point might describe the scaling of membranes which are stationary on average, but where inhomogeneities in the distribution of activators can stimulate membrane growth nonlinearly. 

The physical implications of the arisal of the new fixed point $P_2$ %stable fixed point for the IR scaling of correlation functions 
can be made clear by solving \autoref{eq:RG-equation} with the method of characteristics~\cite{amitFieldTheory,Vasilev04,tauberCriticalDynamics}. The solution shows that, for all parameters $g$ in the basin of attraction of an IR-stable fixed point $g^*$ (such as $P_2$), the following scaling limit holds~\cite{Vasilev04,tauberCriticalDynamics}
%\begin{align}
%C^{\psi_{i}\psi_{j}} %& 
%(k,\omega, g,\nu, m,\mu) \to %\nonumber \\
%%& \to 
%k^{\Delta_{\psi_{i}}^{*}+\Delta_{\psi_{j}}^{*}-d-z^{*}} C\left(\frac{\omega}{\nu k^{z^{*}}}, g^{*}% \right)
%\end{align}
\begin{equation}
C^{\psi_{i}\psi_{j}} \sim %%& 
%(k,\omega, g,\nu, m,\mu) \to %\nonumber \\
%& \to 
k^{y_{\psi_{i}}^{*}+y_{\psi_{j}}^{*}-d-z^{*}} C\left(\frac{\omega}{\nu k^{z^{*}}}, g^{*} \right)
\end{equation}
where
\begin{align}
y_{\psi}^{*} = y_{\psi} + \gamma_{\psi}^{*}, \quad z^{*} = 2 - \gamma_{D}^{*}, \quad \gamma_{\psi}^{*} = \gamma_{\psi}|_{g\rightarrow g^{*}}.
\end{align}
The $y^{*}$'s here are the actual scaling dimensions of the fields, as opposed to the engineering dimensions $y$ which coincide with the $y^{*}$'s only when the nonlinearities are irrelevant. In the present case, from~\autoref{eq:r-c-KPZ-Znu}, \autoref{eq:r-c-KPZ-Zhth} and \autoref{eq:r-c-KPZ-Zptp}, we obtain
\begin{equation}\label{eq:KPZ-wilsonflow}
\gamma_\nu = \frac{\alpha\lambda}{8 r},\, \gamma_h = \frac{\alpha\lambda-\alpha^2-r\lambda^2}{16 r},\, \gamma_\phi = \frac{\lambda(\alpha-\lambda)}{4 (1+r)^2}
\end{equation}
together with $ \gamma_{\tilde{h}} = - \gamma_{h}, \ \gamma_{\tilde{\phi}} = - \gamma_{\phi}$. At the fixed point $P_2$, $\lambda\,{=}\,0$ and $\alpha^2\,{=}\,8r\epsilon$, $\gamma_\nu\,{=}\,\gamma_\phi\,{=}\,0$, whereas $\gamma_h\,{=}\,-\epsilon/2$. Therefore, in the vicinity of $d\,{=}2$, $z^{*}$ and $y_\phi^{*}$ remain at their mean-field values $2$ and $d/2$, respectively. %---typical of normal diffusion for both space-time scaling and density fluctuations. 
The scaling dimension of the height, instead changes from $-\epsilon/2$ in mean-field, to $-\epsilon\,{=}\,d-2$,
\begin{equation}
 y_h = \frac{d-2}{2} \to y_h + \gamma_h^* = d-2.
\end{equation}
Below $d\,{=}\,2$ the scaling dimension of the height is therefore reduced, hence the roughness exponent of the interface %\eqref{eq:Width} 
is increased. Extrapolating the perturbative result to $d\,{=}\,1$, for instance, gives a roughness exponent of $1$, signalling an interface that is rougher than in the mean-field approximation, and whose fluctuations are not those of a standard KPZ interface. Although this result holds at one-loop only, it suggests that the scaling of height fluctuations in the active KPZ model at $\lambda=0$ follows a new behavior, distinct from those shown by KPZ interfaces. 

We close this section with some comments about possible strong-coupling behavior. It is clear from the RG flows depicted in Fig. \ref{fig:phase-diagram-KPZ}, that a run-away solution exists for both $ d\,{<}\, 2 $, and $ d\,{\geq}\, 2$ (though there is a region of attraction to the Gaussian fixed point in the latter case). Based on this feature, we suspect that the non-perturbative fixed point, the exact form of which cannot be captured by our perturbative analysis, may represent  a new novel universality class with $ \lambda^{*},\alpha^{*} \neq 0 $, distinct from the standard KPZ universality class~\cite{Canet2010,Canet2011}. Further non-perturbative analysis is required to elucidate the properties of this fixed point.

\section{Renormalization of other diffusive regimes}\label{sec:renormalization-diffusive-2}

While \autoref{sec:renormalization-diffusive} has dealt with one of the diffusive regimes, two more remain, for which our power counting analysis suggests that a full renormalization group analysis is required. In both these cases (corresponding to the curvotactic activators and passive sliders models), the one-loop renormalization procedure is similar to the one reported above for the active KPZ model. As the algebra is more cumbersome, we refer to~Appendix C %\autoref{app:perturbationCURVO} 
for most of the calculations, and discuss the results in this section.

\subsection{Renormalization of the curvotactic activators model}\label{ssec:curvotactic-renormalization}

We first consider renormalization of the curvotactic activators model, corresponding to an active membrane with $a_\phi\,{=}\,0$ (orange line in the phase diagram in \autoref{fig:phase-diagram}). The equations to consider are
\begin{subequations}\label{eq:active-interface-diffusive-curvotactic}
\begin{align}\label{eq:a-i-diff-i-curvotactic}
\partial_t h &=  q\nu\left(\phi  + \partial_a^2 h\right) +\frac{\nu\lambda}{2}\left(\partial_a h\right)^2,\\ 
\label{eq:a-i-diff-dens-curvotactic}
\partial_t \phi &= - \nu c \partial_a^4 h + \nu \partial_a^2 \phi+ \partial_a\left(\sqrt{2 \nu}\xi_a\right) \nonumber\\ &+ \nu\lambda\partial_a\left(\phi\partial_a h\right) +\frac{\nu\kappa}{2}\left[\partial_a^2 (\partial_b h)^2 -\partial_a\left((\partial_a h)\partial_b^2 h\right)\right],
\end{align}
\end{subequations}
where we have set $\nu_h\,{=}\,q\nu$ and $\nu_\phi\,{=}\nu$, with $q$ the inverse of the parameter $r$ of the previous section. We have also rescaled the coefficients $c$, $\lambda$ and $\kappa$ with $\nu$. As discussed in~\autoref{ssec:curvotactic}, the upper critical dimension of this model is $d_c=4$. Recalling \autoref{eq:primitive-degree-divergence-diffusive} for the primitive degree of divergence of perturbative corrections, and using the engineering dimensions of \autoref{tab:engineering-dimensions-activeh} with $d=d_c=4$,
\begin{equation}\label{eq:diagram-prim-div-deth}
 \delta(\mathcal{I}) = 6 - 4n_{\ti{h}} - 2 n_\phi -2 n_{\ti{\phi}}.
\end{equation}
Taking into account that $ h $ and $ \phi' $ fields must be coupled with gradients, the only diverging vertex functions at $d\,{=}\,4$ are found to be
\begin{equation}\label{eq:vertex-functions-activeh}\begin{aligned}
 \Gamma^{\ti{h}h},\quad \Gamma^{\ti{h}\phi},\quad \Gamma^{\ti{\phi}\ti{\phi}},\quad \Gamma^{\ti{\phi}\phi},\quad \Gamma^{\ti{\phi}h},\quad \\ \Gamma^{\ti{h}hh},\quad \Gamma^{\ti{\phi}\phi h},\quad \Gamma^{\ti{\phi}hh},\quad \Gamma^{\ti{\phi}hhh}.
\end{aligned}\end{equation}
\autoref{eq:vertex-functions-activeh} suggests the addition of two new nonlinear vertices to the action, one proportional to $\ti{\phi}hh$ and one to $\ti{\phi}hhh$. The former simply corresponds to the $\kappa$ term in the equations of motion, \autoref{eq:active-interface-fluctuating-full}, the latter is here excluded because it violates the symmetry for infinitesimal tilt transformations.

One-loop corrections around $d\,{=}\,4$ are computed in~Appendix C. %\autoref{app:perturbationCURVO}. 
The ensuing beta functions, also shown in~Appendix C,%\autoref{app:perturbationCURVO}, 
display two fixed points (besides the mean-field one) in the $q-c-\lambda-\kappa$ parameter space. The fixed points of the RG flow are summarised in~\autoref{tab:f-p-curvotactic}, together with the eigenvalues of the linearised RG flow.
\begin{table}[ht!]
	\def\arraystretch{1.3}
	\centering
	\begin{tabular}{| c || c | c | c | c || c | c | c | c |}
		\hline \hline
		$\,$ & $\quad q\quad$ & $\quad c\quad$ &  $\lambda$ & $\kappa$ & $e_1$ & $e_2$ & $e_3$ & $e_4$ \\  \hline
		$P_0 $ & $\text{any}$ & $\text{any}$ & $0$ & $0$ & $0$ & $0$ & $-\epsilon/2$ & $-\epsilon/2$ \\ \hline
		$P_1$ & $0$ & $0$ & $\pm\sqrt{2\epsilon}$ & $0$ & $-\frac{\epsilon}{2}$ & $\frac{\epsilon}{2}$ & $\frac{17\epsilon}{24}$ & $\epsilon$ \\ \hline
		$P_2$ & $0.5(8)$ & $-3.5(0)$ & $\pm 6.8(4)\sqrt{\epsilon}$ & $0$ & $-$ & $-$ & $-$ & $-$ \\ \hline
	\end{tabular}
	\caption{Fixed points of the RG flow, and corresponding eigenvalues, for the curvotactic activators model. The eigenvalues of the fixed point $P_2$ cannot be written in explicit form for general $\epsilon$. However, as $P_2$ lies outside of the stability region of the parameter space, the flow around $P_2$ does not have physical meaning (hence we do not show the corresponding eigenvalues).}
	\label{tab:f-p-curvotactic}
\end{table}

First, we note the presence of the mean-field fixed point $P_0$, which is repulsive (in the $\lambda$ and $\kappa$ directions)for $d\,{<}\,4$ and attractive for $d\,{>}\,4$. Regarding $P_2$, we recall that, with the present choice of parameters, the stability condition of \autoref{eq:stability-condition} implies $1+c\,{>}\,0$. Therefore, the fixed point $P_2$ lies outside the stability region of the phase diagram. This fixed point is also repulsive, when the flow is projected on the $q-c$ plane, thus it does not have any particular implication for the properties of the system. $P_1$, on the other hand, requires a more detailed analysis.

The sign of the stability matrix eigenvalues at $P_1$ denotes the presence of one unstable direction and three stable ones. The second eigenvector, $\bm{e}_2 \,{=}\,\left(0,1/\sqrt{2\epsilon},2,2\right)$, is the only one having a component in the $\kappa$ direction. As the corresponding eigenvalue $e_2\,{=}\,\epsilon/2$ is positive below $d_c$, we conclude that the flow around $P_1$ is attractive in the $\kappa$ direction. The other two eigenvectors with positive eigenvalue are $\bm{e}_3 \,{=}\,\left(0,7/24\sqrt{2\epsilon},1,0\right)$ and $\bm{e}_4 \,{=}\,\left(0,0,1,0\right)\,{=}\,\bm{e}_\lambda$, signalling an attractive flow in the $c-\lambda$ plane around the fixed point. However, the last eigenvector $\bm{e}_1\,{=}\,\left(3\sqrt{2}/\sqrt{\epsilon},-3\sqrt{2}/\sqrt{\epsilon},1,0\right)$, whose eigenvalue $e_1$ is negative below $d\,{=}\,4$, has a component in the $\lambda$ direction too, implying that the coupling $\lambda$ ultimately flows away towards non-perturbative values. Additionally, we expect that any physical realisation of the equations of motion in~\autoref{eq:active-interface-diffusive-curvotactic} will always have $r>0$. This suggests that the scaling laws in the curvotactic activators model, for $d<d_c$ and for parameter values which avoid linear instabilities, are controlled by a non--perturbative fixed point.
\begin{figure}[t!]
\begin{center}
  \begin{tabular}{cc}
       \includegraphics[width=1.0\columnwidth]{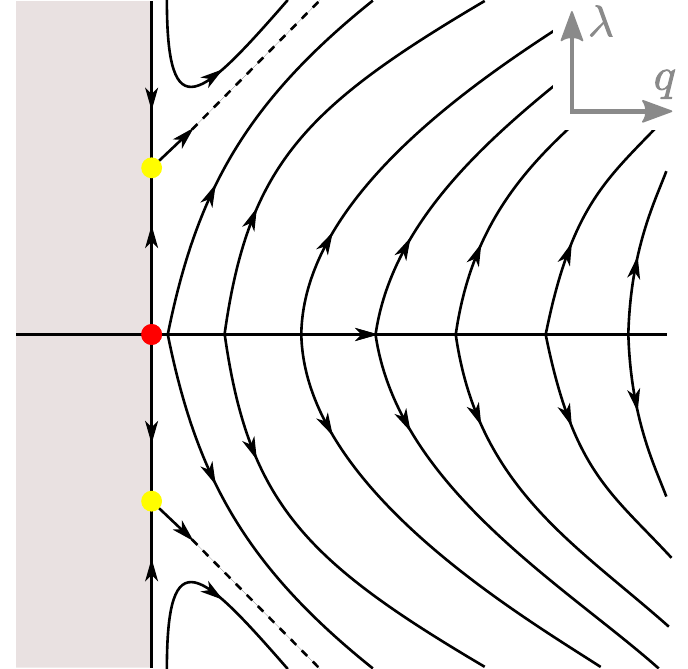}\\
  \end{tabular}
\caption{Phase diagram of the curvotactic activators model, for $d\,{<}\,4$, in the $q-\lambda$ plane ($q$ on the x-axis, $\lambda$ on the y-axis). It should be noted that, as soon as $q\neq 0$, $c$ and $\kappa$ fly away from the respective fixed-point values together with $\lambda$.}
\label{fig:phase-diagram-CURVO}
\end{center}
\end{figure}

\subsection{Renormalization of the passive sliders model}\label{ssec:passivesliders-renormalization}

We now turn to the passive sliders model, corresponding to the $a_h\,{=}\,0$ line of the phase diagram (blue line in \autoref{fig:phase-diagram}). Here, the equations to consider are:
\begin{subequations}\label{eq:active-interface-diffusive-passivesliders}
\begin{align}\label{eq:a-i-diff-i-passivesliders}
\partial_t h &= -c_h\partial_a^2 \phi + \nu_h\partial_a^2 h + \frac{\lambda}{2}\left(\partial_a h\right)^2+\sqrt{2 D_h}\xi_n,\\ 
\label{eq:a-i-diff-dens-passivesliders}
\partial_t \phi &= a_\phi\partial_a^2 h + \nu_\phi \partial_a^2 \phi+ \lambda\partial_a\left(\phi\partial_a h\right).
\end{align}
\end{subequations}
Below $d\,{=}\,2$ the nonlinear terms should be taken into account with a renormalization procedure. However, in this case the model turns out to be non-renormalizable, as there are an infinite number of marginal terms that require renormalization.

Recalling \autoref{eq:primitive-degree-divergence-diffusive} for the primitive degree of divergence of perturbative corrections, using the engineering dimensions of \autoref{tab:engineering-dimensions-passivep} and $d\,{=}\,2$,
\begin{equation}\label{prim-deg-div-passivep}
 \delta(\mathcal{I}) = 4 - 2n_{\ti{h}} -2 n_{\ti{\phi}}.
\end{equation}
The crucial difference between the passive slider model and previously considered models comes from the absence of the number of density fields $ \phi $ in \eqref{prim-deg-div-passivep}. Although the fields $ h $ and $ \phi' $ are still coupled with gradient operators, this does not apply to $ \phi $. Divergent vertex functions at $d\,{=}\,2$ are then found to be
\begin{equation}\label{eq:vertex-functions-pasivep}\begin{aligned}
\Gamma^{\ti{h}\phi^{i}}, \quad \Gamma^{\ti{h}h\phi^j},\quad
\Gamma^{\ti{h}hh\phi^k} , \quad \Gamma^{\ti{h}\ti{h}\phi^l},\quad
\Gamma^{\ti{\phi}\phi^m},\quad \Gamma^{\ti{\phi}h\phi^n},
\end{aligned}\end{equation}
for any non-negative value of $i,j,k,l,m$ and $n$. There are then $6$ infinite families of terms that would need to be included in the theory in order to perform renormalization around $d\,{=}\,2$. This is caused by the engineering dimension of the field $\phi$ which, by vanishing at $d\,{=}\,2$, implies a breakdown of the small $\phi$ approximation used in deriving the model. 

We note that similar models, but without the curvature term proportional to $c_h$, have been considered before~\cite{das2000,drossel2002}, and have been shown to be associated with ``fluctuation-dominated phase ordering'', where the system evolves to a state with long-range order and macroscopic fluctuations. While such large scale fluctuations are in line with our power counting results in \autoref{ssec:passivesliders}, no definite scaling picture has yet been found in $d\,{=}\,1$, which may be due to the complex RG picture we find here. Indeed, it was shown that non-renormalizability may be associated with non-universal behavior~\cite{antonov2017}. However, the special case of \cite{das2000,drossel2002} deserves further analysis from the RG point of view, as tuning $c_h\,{=}\,0$ eliminates most of the diagrams in 
\autoref{eq:vertex-functions-pasivep}. 

\section{Discussion and conclusions}\label{sec:conclusions}

In this work we have introduced  a general continuum model of the plasma membrane and studied its critical properties with field-theoretical techniques. The model includes a field for the height of the membrane, which is assumed to have no overhangs, and a field for the density of ``activators''. The membrane performs overdamped motion vertically, within the environment, and the activators perform overdamped motion horizontally, within the membrane. A coupling between the density of activators and the local membrane slope arises kinematically because of the membrane motion. We performed a detailed RG analysis of the model, which has allowed us to identify four different scaling regimes, where the membrane and activators dynamics are described by a different pair of equations for membrane height and density of activators. The different regimes we unveil encompass and generalise active membrane models which were introduced previously in order to describe specific applications~\cite{prost1996,ramaswamy2000,gov2006,veksler2007,cagnetta1,cagnetta2,gov2005,maitra2014,prost1998}.

We have shown that the scale-invariant properties of an active membrane are determined essentially by two parameters (see~\autoref{fig:phase-diagram}). One is the aforementioned kinematic coupling between the activator density and the membrane slope, which we call $a_\phi$ and is proportional to the average density of activators and the average vertical velocity of the membrane (we called this velocity $\lambda$ throughout the paper). The other key parameter is $a_h$, which quantifies the strength of activator-induced interfacial growth. The first universality class we identify corresponds to the case where $a_\phi a_h>0$---we call this the ``generic active membrane model'' because it is found for any advancing membrane whose growth is stimulated by activators. This case is typical of  the lamellipodium---a supramembrane structure formed at the leading edge of eukaryotic cells which has attracted much attention in biophysics, especially with respect to the observation of ubiquitous lateral waves travelling along the leading edge. Indeed we find that the natural scaling regime for this model is acoustic, i.e. the dispersion relation which links mode frequency and wavevector is linear in the modulus of the wavevector when the latter is small. The wave speed emerging from the dispersion relation is proportional to the speed of the vertical motion of the membrane, thus establishing a strong and testable link between cell motility and lateral waves for generic active membranes.

Our RG analysis shows mean-field theory to hold exactly in all spatial dimension for the generic active membrane model. This implies that the scaling dimensions computed by naive power counting coincide with the exact scaling dimensions of the fields, and, for early times, there are no corrections to the dynamic exponent $z\,{=}\,1$  due to the nonlinear couplings of the model. However, viscous terms in the height (surface tension) and density (diffusion) equations are dangerously irrelevant, and if absent preclude the existence of a steady state. In practice, this means that small non-zero viscous terms have to be included for the theory to behave well in the $t\to \infty$ limit, and, in the presence of these terms, the acoustic regime is relevant up to times of order of the system size $L$. The early-time dynamics in the acoustic regime leaves a detectable signature in the structure factors of membrane and activator density fluctations. %both the membrane width dynamics, which features oscillations superposed on a standard Edwards-Wilkinson behavior~\cite{cagnetta1}, and in the activator density distribution, which is hyperuniform at times $t\sim L$. %The reason for the peculiar acoustic scaling is the interaction with the activator density, which is itself coupled to the gradients of the height because of kinematic coupling between activators and membrane slope.

Three additional scaling regimes emerge when $a_\phi a_h=0$ (these lie on the axes of~\autoref{fig:phase-diagram}), which can be realised either by having an active force which is nonlinear in the activator density, or by keeping the membrane stationary, for instance by applying a suitable external force which exactly balances the average active force. In all these cases, the mean-field dynamical exponent is $z=2$, corresponding to a diffusive scaling of length and timescales. If both $a_h$ and $a_\phi$ are equal to $0$, for instance, the system is described by what we call the ``active KPZ model''.  In this case the naive upper critical dimension of the model, beyond which mean-field theory works exactly (from the perturbative point of view), is $2$.  In the active KPZ model, power counting shows that an additional nonlinear coupling must be included in the theory---the  term $\alpha\phi^2/2$---which describes catalytic, or cooperative, membrane growth: such a term can arise, for instance, when activators are dilute and require dimerisation to stimulate growth. The other relevant nonlinearity in this case is the usual KPZ-term proportional to the squared slope with coefficient $\lambda$. When $\lambda=0$, our one-loop RG calculation shows the emergence of a nontrivial perturbative fixed point controlling the scaling of the active KPZ model for $d<d_c$. This fixed point corresponds to a rough interface, with a larger roughness exponent with respect to that of KPZ and other  passive interfaces.

When $a_\phi\,{=}\,0$, but $a_h\ne 0$, which corresponds to the case of a membrane which is stationary on average, our analysis shows that coupling between membrane curvature and activator density becomes marginal and needs to be included in the model---such a term was, by contrast, irrelevant when $a_h,a_\phi \ne 0$. We call the equations describing this scaling regime the `curvotactic activators model', because the dynamics of the activators is influenced by the interface curvature. A simplified linear version of this model was introduced originally in~\cite{gov2006}, where it was also shown to display transient transverse waves. As the dynamical scaling of this model is diffusive, our analysis shows that the waves seen in this regime are fundamentally different from those which are found in the generic active membrane model, and it would be of interest to perform targeted experiments in cellular systems to find out which of these more closely represents the lateral actin waves found in cells. From a scaling perspective, this model shows enhanced height fluctuations in mean field, leading to a ``super-rough'' scaling with the roughness exponent $\chi \geq 1$ in the physically relevant dimensions $d\,{=}\,1$ and $d\,{=}\,2$. This is most likely due to the noise acting on the density field transferring to the height field through the active-force term $a_h\phi$. Our RG calculations show that the naive upper critical dimension of the curvotactic activators model is $d_c=4$, and that below $d_c$ nonlinearities take the system away from the mean field Gaussian fixed point, most likely towards a strong-coupling fixed point associated with non-perturbative values of both the KPZ parameter $\lambda$ and the new parameter $\kappa$ which emerged from power counting. It would thus be of interest to study this case further with non-perturbative approahes or numerical simulations~\cite{alias2020}.

Finally, when $a_h=0$, the model describes an ensemble of particles that are advected by the slopes on an advancing membrane. These could be, for instance, proteins which lie at the leading edge of a cell without stimulating its growth. This case corresponds to our final scaling regime, that of the ``passive sliders model'', which was studied before in the absence of any curvature coupling, in~\cite{das2000,drossel2002,nagar2005}. This problem is known to show ``fluctuation-dominated phase ordering'', a phenomenon  associated with anomalous giant fluctuations in the density. A power counting analysis indeed reveals  a scaling of density fluctuations  larger than that of a diffusing field,  again due to a noise-transfer effect---this time from the height equation into the density equation through the $a_h$ coupling. However, the model is non-renormalizable below the upper critical dimension, which  here is $d_c\,{=}\,2$. This fact may be at the origin of the highly nontrivial scaling of density fluctuations found numerically in the passive sliders model in $d\,{=}\,1$, and hints at the possibility that the behavior seen there may even be non-universal.
It is interesting to note, in this respect, that tuning $c_h\,{=}\,0$, which corresponds to the case considered in~\cite{das2000,drossel2002,nagar2005}, eliminates all one-loop corrections to the vertex functions in ~\autoref{eq:vertex-functions-pasivep}, which are responsible for non-renormalizability of the model. In other words, when $c_h\,{=}\,0$, none of the remaining parameters in ~\autoref{eq:vertex-functions-pasivep} affects the functions in~\autoref{eq:vertex-functions-pasivep} at the one-loop level, so that the latter remain zero under rescaling if they are not included in the original theory. If this remains true beyond one-loop, the model with $c_h=0$ would become renormalizable again. It would therefore be of considerable interest to study the $c_h=0$ limit by large-scale simulations and non-perturbative approaches in different dimensions to understand in depth the corresponding physics.

We close by noting that, while we have not studied here the case of $a_h a_\phi\,{<}\,0$, because it leads to the breakdown of our approximation of nearly-flat membranes required for the Monge gauge description, this case is far from being uninteresting and certainly deserves further attention in the future. Numerical studies of the unstable phase indeed show a rich phenomenology, e.g. the formation of motile protrusions~\cite{kabaso2011theoretical, fosnaric2019theoretical, sadhu2021modelling}. We shall also mention here that a similar picture, with an unstable-to-stable transition, was found in a two-species lattice model~\cite{ramaswamy2002} inspired by the physics of sedimenting colloidal crystals~\cite{lahiri1997,lahiri2000,chajwa2020}. In the stable phase, corresponding to our generic active membrane model, fluctuations of the fields in a reference frame co-moving with the lateral waves are found to exhibit a wealth of different dynamic scaling regimes~\cite{das2001,chakraborty2019}. In the unstable phase, corresponding to the $a_h a_\phi\,{<}\,0$ sector of our model, different phases are found with a different degree of long-range order in the system~\cite{chakraborty2016,chakraborty2017a,chakraborty2017b, mahapatra2020}. It would be of interest to work out the relevance of the unstable phase for active membrane physics, and then understand the relationship between these one-dimensional results and the picture emerging from our perturbative RG approach.

%Therefore, our study has revealed that  remarkably  diverse physics emerges from the equations \ref{eq:active-interface-fluctuating-full},  which are dictated by general considerations. These equations unify previously studied cases but also have revealed new fixed points that demand further investigation.

\section*{ACKNOWLEDGEMENTS}

FC and V\v{S} contributed equally to this work. The authors would like to thank Juha Honkonen, Ananyo Maitra and Sriram Ramaswamy for many illuminating discussions. FC and DM acknowledge funding from ERC (Consolidator Grant THREEDCELLPHYSICS, 648050). V\v{S} acknowledges funding from EPSRC Grant No. EP/L015110/1 under a studentship.

\appendix

\section{Perturbation theory and Feynman rules}\label{app:feynman}

In this Appendix we discuss the details of the perturbation theory which led to the corrections shown in~\hyperref[eq:KPZ-Ghth-oneloop]{Eqs.~(\ref*{eq:KPZ-Ghth-oneloop}-\ref*{eq:KPZ-Ghtpp-oneloop})}. The actual corrections are computed in the following Appendices. Here we begin by detailing the Fourier-space expression of the harmonic action, \autoref{eq:gaussian-action}. For a generic field $\psi(\bm{x},t)$, we define the space-time Fourier transform as
\begin{equation}\label{eq:fourier}
\psi(\bm{x},t) = \int \frac{d^d k}{(2\pi)^d}\frac{d\omega}{(2\pi)}\psi(\bm{k},\omega)e^{i\left(\bm{k}\cdot\bm{x}-\omega t\right)}.
\end{equation}
We will, in the following, use the shorthand $\Bbbk$ to represent both frequency and momentum, and $\int_\Bbbk$ for the $d\,{+}\,1$-dimensional momentum and frequency integral with infinite-volume normalisation $(2\pi)^{d+1}$. Therefore, \autoref{eq:gaussian-action} can be written as
\begin{widetext}\begin{equation}\label{eq:gaussian-action-exp}
S_0[\ti{h},h,\ti{\phi},\phi] =\frac{1}{2}\int_\Bbbk \left(\begin{array}{c} \tilde{h}(-\Bbbk) \\ {h}(-\Bbbk) \\ \tilde{\phi}(-\Bbbk) \\ {\phi}(-\Bbbk) \end{array}\right)^T \left(\begin{array}{cccc}
 -2D_h & L_h & 0 & -(a_h+c_h k^2) \\
 L_h^\dag & 0 & +k^2 (a_\phi+c_\phi k^2) & 0 \\
 0 & +k^2(a_\phi+c_\phi k^2) & - 2 D_\phi k^2  & L_\phi \\
 -(a_h+c_h k^2) & 0 & L_\phi^\dag & 0 \\
 \end{array}\right) \left(\begin{array}{c} \tilde{h}(\Bbbk) \\ {h}(\Bbbk) \\ \tilde{\phi}(\Bbbk) \\ {\phi}(\Bbbk) \end{array}\right),
\end{equation}\end{widetext}
where we have also introduced the linear operators $L_h\equiv -i\omega +\nu_h k^2$, $L_\phi\equiv -i\omega + \nu_\phi k^2$ and their complex conjugates $L_h^\dag$ and $L_\phi^\dag$. The superscript $^\dag$ denotes transposition.

In complete analogy with \autoref{eq:path-probability}, the harmonic action $S_0$ defines a path probability for the fields $h$ and $\phi$. This harmonic path probability is equivalent to the linearised version of the equations of motion. Inverting the matrix which appears in \autoref{eq:gaussian-action-exp} yields, as in \autoref{eq:gaussian-correlations}, the correlations of the linearised problem,
\begin{subequations}\label{eq:gaussian-two-point}
\begin{align}
 \label{eq:g-c-htht} C^{\ti{h}\ti{h}}_0(\Bbbk) &= 0;\\
 \label{eq:g-c-hth} C^{\ti{h}h}_0(\Bbbk) &= \frac{L^\dag_\phi}{L_h^\dag L_\phi^\dag + k^2\left(a_h + k^2 c_h\right)\left(a_\phi+k^2 c_\phi \right)};\\
 \label{eq:g-c-htpt} C^{\ti{h}\ti{\phi}}_0(\Bbbk) &=0;\\
 \label{eq:g-c-htp} C^{\ti{h}\phi}_0(\Bbbk) &= -\frac{k^2 \left(a_\phi+k^2 c_\phi\right)}{L_h^\dag L_\phi^\dag + k^2\left(a_h + k^2 c_h\right)\left(a_\phi+k^2 c_\phi \right)};\\
 \label{eq:g-c-hht} C^{h\ti{h}}_0(\Bbbk) &= C^{\ti{h}h}_0(-\Bbbk);\\
 \label{eq:g-c-hh} C^{hh}_0(\Bbbk) &= \frac{2D_h|L_\phi|^2 + 2 D_\phi k^2\left(a_h + k^2 c_h\right)^2}{|L_h L_\phi + k^2\left(a_h + k^2 c_h\right)\left(a_\phi+k^2 c_\phi \right)|^2};\\
 \label{eq:g-c-hpt} C^{h\ti{\phi}}_0(\Bbbk) &= \frac{a_h+k^2 c_h}{L_h L_\phi + k^2\left(a_h + k^2 c_h\right)\left(a_\phi+k^2 c_\phi \right)};\\
  \label{eq:g-c-hp} C^{h\phi}_0(\Bbbk) &= \frac{2D_\phi k^2\left(a_h+k^2 c_h\right)L_h^\dag}{|L_h L_\phi + k^2\left(a_h + k^2 c_h\right)\left(a_\phi+k^2 c_\phi \right)|^2} + \nonumber\\ 
  &-\frac{2D_h k^2 \left(a_\phi+k^2 c_\phi\right)L_\phi}{|L_h L_\phi + k^2\left(a_h + k^2 c_h\right)\left(a_\phi+k^2 c_\phi \right)|^2};\\
 \label{eq:g-c-ptht} C^{\ti{\phi}\ti{h}}_0(\Bbbk) &= 0;\\
 \label{eq:g-c-pth} C^{\ti{\phi}h}_0(\Bbbk) &= C^{h\ti{\phi}}_0(-\Bbbk);\\
 \label{eq:g-c-ptpt} C^{\ti{\phi}\ti{\phi}}_0(\Bbbk) &= 0;\\
 \label{eq:g-c-ptp} C^{\ti{\phi}\phi}_0(\Bbbk) &= \frac{L_h^\dag}{L_h^\dag L_\phi^\dag + k^2\left(a_h + k^2 c_h\right)\left(a_\phi+k^2 c_\phi \right)};\\
  \label{eq:g-c-pht} C^{\phi\ti{h}}_0(\Bbbk) &= C^{\ti{h}\phi}_0(-\Bbbk);\\
 \label{eq:g-c-ph} C^{\phi h}_0(\Bbbk) &= C^{h\phi}_0(-\Bbbk);\\
 \label{eq:g-c-ppt} C^{\phi\ti{\phi}}_0(\Bbbk) &= C^{\ti{\phi}\phi}_0(-\Bbbk);\\
 \label{eq:g-c-pp} C^{\phi\phi}_0(\Bbbk) &= \frac{2D_\phi k^2|L_h|^2+ 2D_h k^4\left(a_\phi+k^2 c_\phi\right)^2}{|L_h L_\phi + k^2\left(a_h + k^2 c_h\right)\left(a_\phi+k^2 c_\phi \right)|^2}.
\end{align}
\end{subequations}

The remaining part of the action comprises, in the most general case, four additional terms (see \autoref{eq:action}). The first two, proportional to $\lambda$, have the following Fourier-space expression,
\begin{align}\label{eq:action-lambda}
S_\lambda &= \lambda \int_{\Bbbk_0,\Bbbk_1,\Bbbk_2}\left[(\bm{k}_1\cdot\bm{k}_2)\ti{h}(\Bbbk_0)\frac{h(\Bbbk_1)h(\Bbbk_2)}{2} \right.\nonumber\\  &\left. -(\bm{k}_0\cdot\bm{k}_2)\ti{\phi}(\Bbbk_0)\phi(\Bbbk_1)h(\Bbbk_2)\right]\delta(\Bbbk_0+\Bbbk_1+\Bbbk_2).
\end{align}
The third, proportional to $\alpha$, reads
\begin{align}\label{eq:action-alpha}
S_\alpha = -\alpha \int_{\Bbbk_0,\Bbbk_1,\Bbbk_2} \ti{h}(\Bbbk_0)\frac{\phi(\Bbbk_1)\phi(\Bbbk_2)}{2}\delta(\Bbbk_0+\Bbbk_1+\Bbbk_2),    
\end{align}
while the fourth, proportional to $\kappa$, is given by
\begin{align}\label{eq:action-kappa}
S_\kappa &= -\kappa\int_{\Bbbk_0,\Bbbk_1,\Bbbk_2}\left(k_0^2 \bm{k}_1\cdot\bm{k}_2+k_1^2 \bm{k}_0\cdot\bm{k}_2+k_2^2 \bm{k}_0\cdot\bm{k}_1 \right) \nonumber \\ &\times\ti{\phi}(\Bbbk_0)\frac{h(\Bbbk_1)h(\Bbbk_2)}{2}\delta(\Bbbk_0+\Bbbk_1+\Bbbk_2).
\end{align}

Correlation functions of the complete model are defined as averages of some product of fields over the full path probability $e^{-S}$, which we denote with $\avg{.}$. As $S\,{=}\,S_0\,{+}\,S_\lambda\,{+}\,S_\alpha\,{+}\,S_\kappa$, the same correlation function can be written as the average of the same product of fields, multiplied by $e^{-S_\lambda-S_\alpha-S_\kappa}$, over the Gaussian path probability $e^{-S_0}$. The latter average is commonly denoted with $\avg{.}_0$. The Taylor-expansion of the exponentials $e^{-S_\lambda}$/$e^{-S_\alpha}$/$e^{-S_\kappa}$ within the average provides a systematic expansion of the model's correlation functions in terms of the nonlinear coupling parameters $\lambda$, $\alpha$ and $\kappa$. The $0$-th order terms of such expansions are given by the correlation functions of the Gaussian model, which can all be computed as sums of products of the two-point functions in \autoref{eq:gaussian-two-point}. Higher-order terms are conveniently organised with the diagrammatic representation introduced by Feynman. Specifically, each of the two-point function in \autoref{eq:gaussian-two-point} is represented with a directed line, such that $h$/$\phi$ fields are associated with solid/dashed lines and response fields $\ti{h}$ and $\ti{\phi}$ have an additional vertical tick, i.e.
\begin{subequations}\label{eq:feynman-lines}
\begin{flalign}\label{eq:f-l-hht}
\qquad\quad &\raisebox{-0.05cm}{\includegraphics[width=2.2cm]{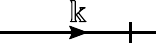}} = C_0^{h\ti{h}}(\Bbbk) = C_0^{\ti{h}h}(-\Bbbk);&
\end{flalign}
\begin{flalign}\label{eq:f-l-hh}
\qquad\quad &\raisebox{-0.05cm}{\includegraphics[width=2.2cm]{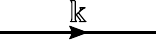}} = C_0^{hh}(\Bbbk);&
\end{flalign}
\begin{flalign}\label{eq:f-l-ppt}
\qquad\quad &\raisebox{-0.05cm}{\includegraphics[width=2.2cm]{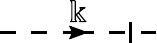}} = C_0^{\phi\ti{\phi}}(\Bbbk) = C_0^{\ti{\phi}\phi}(-\Bbbk);&
\end{flalign}
\begin{flalign}\label{eq:f-l-pp}
\qquad\quad &\raisebox{-0.05cm}{\includegraphics[width=2.2cm]{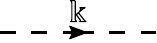}} = C_0^{\phi\phi}(\Bbbk);&
\end{flalign}
\begin{flalign}\label{eq:f-l-hpt}
\qquad\quad &\raisebox{-0.05cm}{\includegraphics[width=2.2cm]{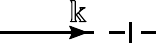}} = C_0^{h\ti{\phi}}(\Bbbk) = C_0^{\ti{\phi}h}(-\Bbbk);&
\end{flalign}
\begin{flalign}\label{eq:f-l-hp}
\qquad\quad &\raisebox{-0.05cm}{\includegraphics[width=2.2cm]{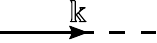}} = C_0^{h\phi}(\Bbbk) = C_0^{\phi h}(-\Bbbk);&
\end{flalign}
\begin{flalign}\label{eq:f-l-pht}
\qquad\quad &\raisebox{-0.05cm}{\includegraphics[width=2.2cm]{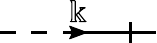}} = C_0^{\phi\ti{h}}(\Bbbk) = C_0^{\ti{h}\phi}(-\Bbbk);&
\end{flalign}
\begin{flalign}\label{eq:f-l-ph}
\qquad\quad &\raisebox{-0.05cm}{\includegraphics[width=2.2cm]{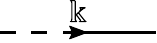}} = C_0^{\phi h}(\Bbbk) = C_0^{h\phi}(-\Bbbk),&
\end{flalign}
\end{subequations}
In addition, the Fourier-space coefficients of nonlinear coupling terms (with a minus sign because $P\propto e^{-S}$) are represented by vertices,
\begin{subequations}\label{eq:feynman-vertices}
\begin{flalign}\label{eq:lambda-vertex-1}
\qquad\quad&\raisebox{-0.7cm}{\includegraphics[width=2.2cm]{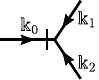}}  = -\frac{\lambda}{2}(\bm{k}_1\cdot\bm{k}_2);&
\end{flalign}
\begin{flalign}\label{eq:lambda-vertex-2}
\qquad\quad&\raisebox{-0.7cm}{\includegraphics[width=2.2cm]{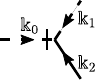}}  = \lambda (\bm{k}_0\cdot\bm{k}_2);&
\end{flalign}
\begin{flalign}\label{eq:alpha-vertex}
\qquad\quad&\raisebox{-0.7cm}{\includegraphics[width=2.2cm]{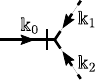}}  = \frac{\alpha}{2};&
\end{flalign}
\begin{flalign}\label{eq:kappa-vertex}
\qquad\quad\raisebox{-0.7cm}{\includegraphics[width=2.2cm]{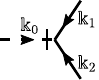}}  &= \frac{\kappa}{2} \left[k_0^2 (\bm{k}_1\cdot\bm{k}_2) + k_1^2 \left(\bm{k}_0\cdot\bm{k}_2\right)\right.\nonumber \\  & \left.+ k_2^2\left(\bm{k}_0\cdot\bm{k}_1\right)\right].& 
\end{flalign}
\end{subequations}
Therefore, the term of order $n$/$m$/$l$ in $\lambda$/$\alpha$/$\kappa$ of a generic $p$-point correlation function $\avg{\prod_{j=1}^p \psi_j(\Bbbk_j)}$ can be obtained as follows. First, one should draw all the topologically distinct diagrams with $p$ ``external'' loose ends (the $j$-th end representing the field $\phi_j$ and carrying momentum $\Bbbk_j$) and $n$ $\lambda$-vertices, $m$ $\alpha$-vertices and $l$ $\kappa$-vertices. Because of the $\delta$-functions in \autoref{eq:action-lambda}, \autoref{eq:action-alpha} and \autoref{eq:action-kappa}, the momenta of incoming and outgoing lines at each vertex should add up to zero. In other words, conservation of momentum must be ensured along the diagram. Second, one should include the ``symmetry factor'' of the diagram, that is a number which includes the $(n!m!l!)^{-1}$ factor coming from the Taylor expansion of the exponential and the multiplicity of the diagram coming from the symmetry of the diagram itself for permutation of vertices and lines. Once the perturbative expansion has been written down in terms of diagrams, corrections can be cast back into functional form by substituting lines according to \autoref{eq:feynman-lines}, vertices according to \autoref{eq:feynman-vertices} and finally performing a $\Bbbk$-integral over the momenta of the ``internal'' lines which are not fixed by the global conservation of momentum.

\section{Perturbative corrections at the $a_h\,{=}\,a_\phi\,{=}\,0$ point}\label{app:perturbationKPZ}

In this Appendix we show, following the rules outlined in Appendix A, %\autoref{app:feynman}, 
that one-loop perturbative corrections to the vertex functions listed in \autoref{eq:divergent-vertices-KPZ} are those presented in \eqref{eq:KPZ-Ghth-oneloop}-\eqref{eq:KPZ-Ghtpp-oneloop}. We shall first recall the general relation between $n$-point connected correlation functions, $C^{\psi_1\dots\psi_n}(\Bbbk_1,\dots,\Bbbk_{n})$, and $n$-point vertex functions, $\Gamma^{\psi_1\dots\psi_n}(\Bbbk_1,\dots,\Bbbk_{n})$. This relation, which can be derived by resorting to the relation \autoref{eq:vertex-gen-fun} between the respective generating functionals, can be compactly written as follows~\cite{amitFieldTheory},
\begin{widetext}\begin{equation}\label{eq:correlations-to-vertices}
    C^{\psi_1\dots\psi_n}(\Bbbk_1,\dots,\Bbbk_{n}) = - \sum_{\varphi_1,\dots,\varphi_n}\left[\prod_{i=1}^n C^{\psi_i \varphi_j}(\Bbbk_i, \Bbbk'_i)\right]\Gamma^{\varphi_1\dots\varphi_n}(\Bbbk'_1,\dots,\Bbbk'_{n}) + \mathcal{Q}^{\psi_1\dots\psi_n}(\Bbbk_1,\dots,\Bbbk_{n}),
\end{equation}\end{widetext}
where the sum over a field $\varphi_i$ denotes the sum over all the fields in the theory, $\ti{h}$, $h$, $\ti{\phi}$ and $\phi$. The function $\mathcal{Q}^{\psi_1\dots\psi_n}(\Bbbk_1,\dots,\Bbbk_{n})$ collects all contributions which are {\it one-particle reducible}. One-particle reducible here means that the corresponding Feynman diagram splits into two distinct non-trivial diagrams upon removing one of the internal lines. As such diagrams are all contained in $\mathcal{Q}$, $\Gamma^{\psi_1\dots\psi_n}(\Bbbk_1,\dots,\Bbbk_{n})$ takes contributions from {\it one-particle irreducible} (1PI) diagrams only, i.e. those which do not separate upon removal of an internal line.

Therefore, \autoref{eq:correlations-to-vertices} provides an operational definition for perturbative corrections to vertex function: consider first the perturbative corrections to a connected correlation function, then discard diagrams which are one-particle reducible by internal cuts, remove the external legs (so as to account for the factor $\prod_{i=1}^n C^{\psi_i \varphi_j}(\Bbbk_i, \Bbbk'_i)$) and finally apply an overall minus sign. Let us consider, for example, $C^{h\ti{h}}(\Bbbk)$. One of the one-loop contributions to $C^{h\ti{h}}(\Bbbk)$ is represented by the following diagram,
\begin{equation}
\raisebox{-0.7cm}{\includegraphics[width=2.8cm]{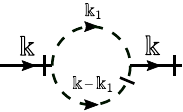}}.
\end{equation}
This diagram is clearly 1PI. Dividing out the two external legs $C_0^{h\ti{h}}(\Bbbk)$ (from the right) and $C^{h\ti{h}}(\Bbbk)=C^{\ti{h}h}(-\Bbbk)$ (from the left) leaves a one-loop correction to $\Gamma^{\ti{h}h}(\Bbbk)$, (notice the shorter external lines)
\begin{equation}\label{eq:Ghth-sample}
\raisebox{-0.7cm}{\includegraphics[width=2.2cm]{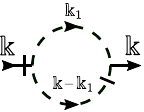}}.
\end{equation}
The contribution of the diagram in \autoref{eq:Ghth-sample} can be computed by following the rules outlined in Appendix A. %\autoref{app:feynman}. 
There is one vertex as in \autoref{eq:alpha-vertex}, one of the kind of \autoref{eq:lambda-vertex-2} and two two-point functions: $C_0^{\phi\ti{\phi}}(\Bbbk-\Bbbk_1)$ and $C_0^{\phi\phi}(\Bbbk_1)$. Exchanging the two $\phi$-lines of the leftmost vertex leaves the diagram unchanged, for a symmetry factor of $2$, which cancels the $1/2$ factor in the $\alpha$-vertex (see \autoref{eq:alpha-vertex}). Hence we get
\begin{widetext}\begin{equation}\begin{aligned}\label{eq:Ghth-sample-calc}
\raisebox{-0.7cm}{\includegraphics[width=2.2cm]{KPZp_GHtH2}} &= \int_{\Bbbk_1}(\nu\alpha)(\nu \lambda)\left[\left(\bm{k}-\bm{k}_1\right)\cdot(-\bm{k})\right]C_0^{\phi\ti{\phi}}(\Bbbk-\Bbbk_1)C_0^{\phi\phi}(\Bbbk_1) \\ &=
-\nu^2 \alpha\lambda \int_{\Bbbk_1} \frac{r\nu k_1^2 \left[\left(\bm{k}-\bm{k}_1\right)\cdot\bm{k}\right]}{\left[\left|-i\omega_1 + r \nu k_1^2\right|^2\right]\left[-i\left(\omega-\omega_1\right) + r\nu \left(\bm{k}-\bm{k}_1\right)^2 \right]},
\end{aligned}\end{equation}\end{widetext}
where, in the second line, $C_0^{\phi\ti{\phi}}(\Bbbk-\Bbbk_1)$ and $C_0^{\phi\phi}(\Bbbk_1)$ have been substituted according to \autoref{eq:g-c-ppt} and \autoref{eq:g-c-pp}, with the model parameters set to the values relevant for the $a_h\,{=}\,a_\phi\,{=}\,0$ point of the phase diagram. Note that counting the powers of momentum in the $\bm{k}_1$-integral, for $d\,{=}\,d_c\,{=}\,2$, yields a primitive degree of divergence of $2$, in agreement with \autoref{eq:primitive-degree-divergence-diffusive}. 

The bare vertex function $\Gamma_0^{\ti{h}h}(\Bbbk)$ coincides with the inverse of $C_0^{h\ti{h}}(\Bbbk)$, that is $\Gamma_0^{\ti{h}h}(\Bbbk)\,{=}\,-i\omega + \nu k^2$. Because the integral on the right-hand side of \autoref{eq:Ghth-sample-calc} vanishes as $\bm{k}\to\bm{0}$, the diagram of \autoref{eq:Ghth-sample} gives no corrections to the $-i\omega$ term of the vertex functions. One can then set $\omega\,{=}\,0$. What remains can be expanded in powers of the external momentum $\bm{k}$: because the primitive degree of divergence of the diagram is $2$, the coefficient of the second-order term of the expansion will display a logarithmic UV divergence, which can be absorbed via a re-definition of the parameter $\nu$. Higher-order terms of the expansion will have coefficients which do not display any UV divergence around $d\,{=}\,2$ and are not relevant for the RG treatment. Therefore, after setting $\omega\,{=}\,0$, performing the $\omega_1$-integral and expanding the result in powers of $\bm{k}$, we get 
\begin{equation}\label{eq:Ghth-sample-calc-exp}
\raisebox{-0.7cm}{\includegraphics[width=2.2cm]{KPZp_GHtH2}} = -\nu k^2 \frac{\alpha\lambda}{4r}\frac{d-1}{d}\int_{\bm{k}_1}\frac{1}{k_1^2} + \mathcal{O}(k^4).
\end{equation}
The $d$-dependent factors encountered in this kind of expression come from the replacement, in the Taylor expansion of the integrand, of the factor $(\bm{k}_1\cdot\bm{k})^2$ with $k_1^2 k^2/d$, which is possible because of the isotropy of the $\bm{k}_1$-integral.

The $\bm{k}_1$-integral on the right-hand side of \autoref{eq:Ghth-sample-calc-exp} displays the expected logarithmic UV divergence in $d\,{=}\,2$. Additionally, it displays an IR divergence for $d\,{\leq}\,2$, primarily because all the relevant parameters of the theory have been set to zero. In order to make the integral finite below $d_c$, an IR regulator must be chosen. The most natural ones at this stage are the relevant parameters $a_h,a_\phi$. This is, however, an inconvenient choice from a practical standpoint, as it makes the structure of propagators much more complex (see \eqref{eq:g-c-htht}-\eqref{eq:g-c-pp}). Other commonly used regulators are finite external momentum \cite{amitFieldTheory} or frequency \cite{frey1994aa}. We consider, instead, a sharp cut-off con the noise correlation, i.e. 
\begin{equation}
\avg{\xi(\bm{k},t)\xi(\bm{k}',t')} \propto \theta(|\bm{k}|-m)\delta(\bm{k}+\bm{k}')\delta(t-t'), 
\end{equation}
where $\theta$ denotes the heavyside step-function and $m$ is an infrared (IR) regulator having the dimensionality of momentum. Such IR-regularisation scheme has been mainly considered in RG studies of the Navier-Stokes equation~\cite{adzhemyan1999field,adzhemyan2003} and turbulent mixing of reaction-diffusion processes \cite{Antonov09,Antonov17,Honkonen18}: the parameter $m$ represents the largest (inverse) lengthscale at which the stochastic noise act. As a result, all the two-point correlation functions with no response fields (\autoref{eq:f-l-hh}, \autoref{eq:f-l-pp}, \autoref{eq:f-l-hp} and \autoref{eq:f-l-ph}) acquire a factor $\theta(|\bm{k}|-m)$. In the case of \autoref{eq:Ghth-sample-calc-exp}, for instance, the $\theta(|\bm{k}_1|-m)$ coming from $C_0^{\phi\phi}(\Bbbk_1)$ cures the IR divergence, allowing us to compute the $\bm{k}_1$-integral for $d\,{<}\,d_c$:
\begin{equation}
\raisebox{-0.7cm}{\includegraphics[width=2.2cm]{KPZp_GHtH2}} = -\nu k^2 \frac{\alpha\lambda}{4r}\frac{d-1}{d}\frac{S_d m^{d-2}}{2-d} + \mathcal{O}(k^4),
\end{equation}
with $S_d$ denoting the measure of the $d$-dimensional hypersphere, scaled by $(2\pi)^d$. Notice that the logarithmic UV divergence for $d\to2$ is now represented by a simple pole $(2-d)^{-1}$.

There is an additional diagram contributing to the one-loop correction of $\Gamma^{\ti{h}h}(\Bbbk)$. However, the coefficient of the $k^2$ term in the Taylor-expansion of such diagram vanishes. To sum up,
\begin{widetext}\begin{equation}\label{eq:KPZ-Ghth-oneloop}
\Gamma^{\ti{h}h}(\Bbbk) = -i\omega + \nu k^2 -
\raisebox{-0.7cm}{\includegraphics[width=2.5cm]{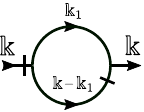}} - \raisebox{-0.7cm}{\includegraphics[width=2.5cm]{KPZp_GHtH2}} = -i\omega + \nu k^2\left[1+\frac{(d-1)\alpha\lambda}{4dr}\frac{S_d m^{d-2}}{2-d}\right] + \mathcal{O}(k^4).
\end{equation}
Analogous considerations yield one-loop corrections for all the vertex functions listed in \autoref{eq:divergent-vertices-KPZ}. For instance,
\begin{equation}\label{eq:KPZ-Ghtht-oneloop}
\Gamma^{\ti{h}\ti{h}}(\Bbbk) = -\nu_h-\frac{1}{2}
\raisebox{-0.7cm}{\includegraphics[width=2.5cm]{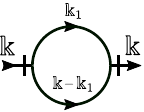}} - \frac{1}{2}\raisebox{-0.7cm}{\includegraphics[width=2.5cm]{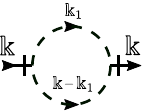}} = -\nu\left[1+\frac{r\lambda^2+\alpha^2}{8r}\frac{S_d m^{d-2}}{2-d}\right] + \mathcal{O}(k^2).
\end{equation}
Note the factor of $1/2$ multiplying the diagrams in \autoref{eq:KPZ-Ghth-oneloop}: the numerical pre-factor for both diagrams is $1/8$ ($1/2$ from each vertex and $1/2$ from the Taylor expansion) , while the symmetry factor is only $4$ (symmetry for exchange of the two vertices and the two internal lines). As the primitive degree of divergence of $\Gamma^{\ti{h}\ti{h}}$ is zero, no Taylor expansion of the integrand of the loop integral was required. Moreover,
\begin{equation}\begin{aligned}\label{eq:KPZ-Gptp-oneloop}
\Gamma^{\ti{\phi}\phi}(\Bbbk) &= -i\omega + r\nu k^2 -
\raisebox{-0.7cm}{\includegraphics[width=2.5cm]{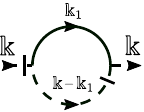}} - \raisebox{-0.7cm}{\includegraphics[width=2.5cm]{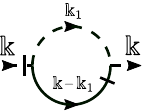}} \\ &= -i\omega + r\nu k^2\left[1+\frac{(1-r)\lambda^2+(d-2+dr)\alpha\lambda}{2dr(1+r)^2}\frac{S_d m^{d-2}}{2-d}\right] + \mathcal{O}(k^4),
\end{aligned}\end{equation}
and
\begin{equation}\label{eq:KPZ-Gptpt-oneloop}
\Gamma^{\ti{\phi}\ti{\phi}}(\Bbbk) = -r\nu k^2 -
\raisebox{-0.7cm}{\includegraphics[width=2.5cm]{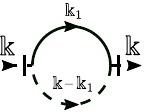}} = = -r\nu k^2\left[ 1+\frac{\lambda^2}{2dr(1+r)}\frac{S_d m^{d-2}}{2-d} \right] + \mathcal{O}(k^4).
\end{equation}

The three point functions $\Gamma^{\ti{h}hh}$ and $\Gamma^{\ti{\phi}\phi h}$ do not receive any correction: the reason is that, because of the infinitesimal tilt symmetry (\autoref{eq:tilt-transformation}), corrections to the $\lambda$-vertices must coincide with the corrections to the $-i\omega$ terms in the two-point vertex functions $\Gamma^{\ti{\phi}\phi}$ and $\Gamma^{\ti{h}h}$, which vanish. One-loop corrections to $\Gamma^{\ti{h}\phi\phi}$, instead, are given by
\begin{equation}\label{eq:KPZ-Ghtpp-oneloop}\begin{aligned}
\Gamma^{\ti{h}\phi\phi}(\Bbbk) = -\alpha\nu &-
\raisebox{-1.3cm}{\includegraphics[width=3.0cm]{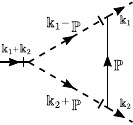}}-
\raisebox{-1.3cm}{\includegraphics[width=3.0cm]{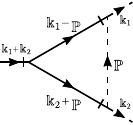}} -\raisebox{-1.3cm}{\includegraphics[width=3.0cm]{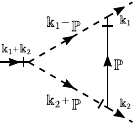}} \\
&-\raisebox{-1.3cm}{\includegraphics[width=3.0cm]{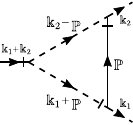}}-\raisebox{-1.3cm}{\includegraphics[width=3.0cm]{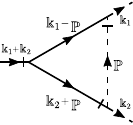}}-
\raisebox{-1.3cm}{\includegraphics[width=3.0cm]{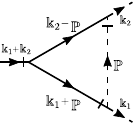}} \\
&= -\alpha\nu\left[1+\frac{(1-r)(\lambda^2-\alpha\lambda)}{2r(1+r)}\frac{S_d m^{d-2}}{2-d}\right] + \mathcal{O}(\bm{k}_1,\bm{k}_2).
\end{aligned}\end{equation}

\section{renormalization on the $a_\phi\,{=}\,0$ line}\label{app:perturbationCURVO}

\newcommand{\LL}{\mathcal{L}}

In this Appendix we give details of the RG analysis on the $a_h$ axis of the reduced phase diagram shown in \autoref{fig:phase-diagram}. Rescaling the parameters in \eqref{eq:gaussian-two-point} as described in Sec. \ref{ssec:curvotactic-renormalization}, the only non-zero propagators have the following form
\begin{subequations}\label{eq:gaussian-two-point-DetH}
	\begin{align}
		%%%%%%%%%%%%%%%%%%%%%%%%%%%%%%%%%%%%%%%%%%%%%%%%%%%%%%%%%%%%%%%%
		C^{\ti{h}h}_0(\Bbbk) &= \frac{L^\dag_\phi}{\LL_{+}^\dag\LL_{-}^\dag}; \quad C^{hh}_0(\Bbbk) = \frac{q^{2} \nu^{3} k^{2}}{|\LL_{+}\LL_{-}|^2}; \quad C^{\ti{\phi}\phi}_0(\Bbbk) = \frac{L_h^\dag}{\LL_{+}^\dag\LL_{-}^\dag}; \quad C^{\phi\phi}_0(\Bbbk) = \frac{\nu k^{2} |L_{h}|^{2}}{|\LL_{+}\LL_{-}|^2}; \\
		%%%%%%%%%%%%%%%%%%%%%%%%%%%%%%%%%%%%%%%%%%%%%%%%%%%%%%%%%%%%%%%%
		C^{h\ti{\phi}}_0(\Bbbk) &= \frac{q \nu}{\LL_{+}\LL_{-}}\ \quad C^{h\phi}_0(\Bbbk) = \frac{q \nu^{2} k^{2} L_{h}^{\dag}}{|\LL_{+}\LL_{-}|^2}; \quad C^{\ti{h}\phi}_0(\Bbbk) = -\frac{\nu c 	k^4}{\LL_{+}^\dag\LL_{-}^\dag};
		%%%%%%%%%%%%%%%%%%%%%%%%%%%%%%%%%%%%%%%%%%%%%%%%%%%%%%%%%%%%%%%%
	\end{align}
\end{subequations}
together with their conjugated counterparts, and we have used the following shorthand notation
\begin{align}
	\begin{aligned}	
		%%%%%%%%%%%%%%%%%%%%%%%%%%%%%%%%%%%%%%%%%%%%%%%%%%%%%%%%%%%%%%%%
		L_{h} &= - i \omega + q \nu k^{2}, \quad L_{\phi} = - i\omega + \nu k^{2}, \\
		%%%%%%%%%%%%%%%%%%%%%%%%%%%%%%%%%%%%%%%%%%%%%%%%%%%%%%%%%%%%%%%%
		\LL_{+} &= - i \omega + \nu k^{2} X_{+}, \quad \LL_{-} = - i \omega + \nu k^{2} X_{-}, \\
		%%%%%%%%%%%%%%%%%%%%%%%%%%%%%%%%%%%%%%%%%%%%%%%%%%%%%%%%%%%%%%%%
		X_{\pm} &= \frac{1}{2} \bigg[ (1+q) \pm \sqrt{(1+q)^{2} - 4q (c+1)} \bigg].
		%%%%%%%%%%%%%%%%%%%%%%%%%%%%%%%%%%%%%%%%%%%%%%%%%%%%%%%%%%%%%%%%
	\end{aligned}
\end{align}
It is worth noting that $ \text{Re}[X_{\pm}] > 0 $, which is in particular useful for evaluating the frequency integrals. The vertex functions that require renormalization are listed in Eq. \eqref{eq:vertex-functions-activeh}. The corresponding perturbative corrections are obtained in a standard fashion

\begin{equation}\label{eq:DetH-GHtH-oneloop}
	\begin{aligned}
		%%%%%%%%%%%%%%%%%%%%%%%%%%%%%%%%%%%%%%%%%%%%%%%%%%%%%%%%%%%%%%%%
		\Gamma^{\tilde{h}h}(\Bbbk) &= -q\nu k^2
		-\raisebox{-0.7cm}{\includegraphics[width=2.5cm]{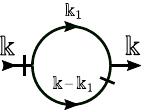}}
		-\raisebox{-0.7cm}{\includegraphics[width=2.5cm]{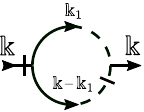}}
		-\raisebox{-0.7cm}{\includegraphics[width=2.5cm]{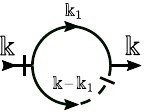}} \\
		%%%%%%%%%%%%%%%%%%%%%%%%%%%%%%%%%%%%%%%%%%%%%%%%%%%%%%%%%%%%%%%%
		&= -q\nu k^2 \left[ 1 + \frac{(d-2)(c+q+2) \lambda^{2} + 2 (d-1) (1+q) \lambda \kappa}{4 (1+c)^2 d (1+q)^2 } \frac{m^{d-4}}{4-d} \right] + \mathcal{O}(k^4).
		%%%%%%%%%%%%%%%%%%%%%%%%%%%%%%%%%%%%%%%%%%%%%%%%%%%%%%%%%%%%%%%%
	\end{aligned}
\end{equation}

\begin{equation}\label{eq:DetH-GHtp-oneloop}
	\begin{aligned}
		%%%%%%%%%%%%%%%%%%%%%%%%%%%%%%%%%%%%%%%%%%%%%%%%%%%%%%%%%%%%%%%%
		\Gamma^{\tilde{h}\phi}(\Bbbk) &= -q\nu 
		-\raisebox{-0.7cm}{\includegraphics[width=2.5cm]{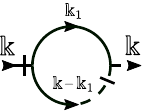}} = -q\nu \left[ 1 - \frac{\lambda^{2}}{4 (1+c)^2 (1+q)} \frac{m^{d-4}}{4-d} \right] + \mathcal{O}(k^2).
		%%%%%%%%%%%%%%%%%%%%%%%%%%%%%%%%%%%%%%%%%%%%%%%%%%%%%%%%%%%%%%%%
	\end{aligned}
\end{equation}
\\

\begin{equation}\label{eq:DetH-GPtPt-oneloop}
	\begin{aligned}
		%%%%%%%%%%%%%%%%%%%%%%%%%%%%%%%%%%%%%%%%%%%%%%%%%%%%%%%%%%%%%%%%
		\Gamma^{\tilde{\phi}\tilde{\phi}}(\Bbbk) &= -\nu k^2
		-\raisebox{-0.7cm}{\includegraphics[width=2.5cm]{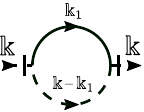}}
		-\raisebox{-0.7cm}{\includegraphics[width=2.5cm]{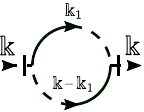}}
		-\frac{1}{2} \raisebox{-0.7cm}{\includegraphics[width=2.5cm]{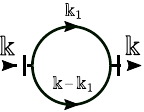}}
		-\raisebox{-0.7cm}{\includegraphics[width=2.5cm]{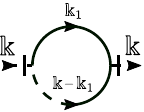}} -\raisebox{-0.7cm}{\includegraphics[width=2.5cm]{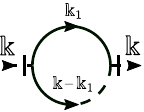}} \\
		%%%%%%%%%%%%%%%%%%%%%%%%%%%%%%%%%%%%%%%%%%%%%%%%%%%%%%%%%%%%%%%%
		&= -\nu k^2 \left[ 1 + \frac{q \lambda^{2}}{2d(1+c)(1+q)^{3}} \frac{m^{d-4}}{4-d} \right] + \mathcal{O}(k^4).
		%%%%%%%%%%%%%%%%%%%%%%%%%%%%%%%%%%%%%%%%%%%%%%%%%%%%%%%%%%%%%%%%
	\end{aligned}
\end{equation}

\begin{equation}\label{eq:DetH-GPtP-oneloop}
	\begin{aligned}
		%%%%%%%%%%%%%%%%%%%%%%%%%%%%%%%%%%%%%%%%%%%%%%%%%%%%%%%%%%%%%%%%
		\Gamma^{\tilde{\phi}\phi}(\Bbbk) &= -\nu k^2
		-\raisebox{-0.7cm}{\includegraphics[width=2.5cm]{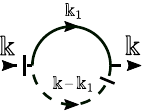}}
		-\raisebox{-0.7cm}{\includegraphics[width=2.5cm]{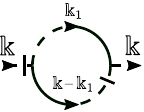}}
		-\raisebox{-0.7cm}{\includegraphics[width=2.5cm]{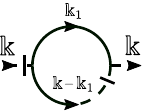}} \\
		%%%%%%%%%%%%%%%%%%%%%%%%%%%%%%%%%%%%%%%%%%%%%%%%%%%%%%%%%%%%%%%%
		&= -r\nu k^2 \left[ 1 + \frac{(c (d-4)-q(d-2)-2) q \lambda^{2}  - 2 (d-1) q (q+1)\lambda \kappa}{4 (1+c)^2 d (1+q)^2 } \frac{m^{4-d}}{4-d} \right] + \mathcal{O}(k^4).
		%%%%%%%%%%%%%%%%%%%%%%%%%%%%%%%%%%%%%%%%%%%%%%%%%%%%%%%%%%%%%%%%
	\end{aligned}
\end{equation}

\begin{equation}\label{eq:DetH-GPtH-oneloop}
	\begin{aligned}
		%%%%%%%%%%%%%%%%%%%%%%%%%%%%%%%%%%%%%%%%%%%%%%%%%%%%%%%%%%%%%%%%
		\Gamma^{\tilde{\phi}h}(\Bbbk) &= c \nu k^4
		-\raisebox{-0.7cm}{\includegraphics[width=2.5cm]{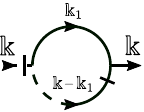}}
		-\raisebox{-0.7cm}{\includegraphics[width=2.5cm]{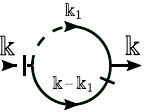}}
		-\raisebox{-0.7cm}{\includegraphics[width=2.5cm]{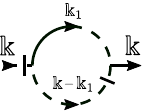}}
		-\raisebox{-0.7cm}{\includegraphics[width=2.5cm]{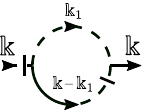}}
		-\raisebox{-0.7cm}{\includegraphics[width=2.5cm]{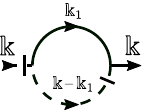}} \\
		%%%%%%%%%%%%%%%%%%%%%%%%%%%%%%%%%%%%%%%%%%%%%%%%%%%%%%%%%%%%%%%%
		&\hspace{1.2cm}
		-\raisebox{-0.7cm}{\includegraphics[width=2.5cm]{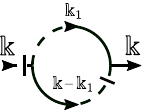}}
		-\raisebox{-0.7cm}{\includegraphics[width=2.5cm]{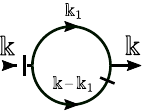}}
		-\raisebox{-0.7cm}{\includegraphics[width=2.5cm]{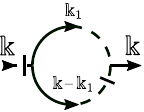}}
		-\raisebox{-0.7cm}{\includegraphics[width=2.5cm]{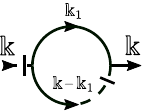}}
		\\
		%%%%%%%%%%%%%%%%%%%%%%%%%%%%%%%%%%%%%%%%%%%%%%%%%%%%%%%%%%%%%%%%
		&= c \nu k^4 \left[ 1 + \frac{A \lambda^{2} + B \lambda \kappa + C \kappa^{2}}{4d(2+d)c(1+c)^{2}(1+q)^{3}}\frac{m^{d-4}}{4-d} \right] + \mathcal{O}(k^6). \\
		%%%%%%%%%%%%%%%%%%%%%%%%%%%%%%%%%%%%%%%%%%%%%%%%%%%%%%%%%%%%%%%%
		A&= -4 c^2 (3 d+2) q-2 c (d q (2 q+15)+d+q (q+10)+1)-q \left(d^2 (q+1)^2-2 d (q-1) (2 q+5)-6 q (q+2)+2\right), \\
		%%%%%%%%%%%%%%%%%%%%%%%%%%%%%%%%%%%%%%%%%%%%%%%%%%%%%%%%%%%%%%%%
		B&= -4 c (d+1) q (q+1)-2 (q+1) \left((d-3) (2 d+3) q^2+2 (d-1) d q-d-8 q-1\right), \\
		%%%%%%%%%%%%%%%%%%%%%%%%%%%%%%%%%%%%%%%%%%%%%%%%%%%%%%%%%%%%%%%%
		C&= -4 \left(d^2-3\right) q (q+1)^2.
		%%%%%%%%%%%%%%%%%%%%%%%%%%%%%%%%%%%%%%%%%%%%%%%%%%%%%%%%%%%%%%%%
	\end{aligned}
\end{equation}

\begin{equation}\label{eq:DetH-GPtHH-oneloop}
	\begin{aligned}
		%%%%%%%%%%%%%%%%%%%%%%%%%%%%%%%%%%%%%%%%%%%%%%%%%%%%%%%%%%%%%%%%
		\Gamma^{\ti{h}\phi\phi}(\Bbbk) = -\alpha\nu &
		-\raisebox{-1.3cm}{\includegraphics[width=3.0cm]{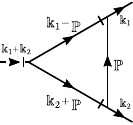}}
		-2 \raisebox{-1.3cm}{\includegraphics[width=3.0cm]{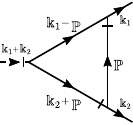}} 
		- \raisebox{-1.3cm}{\includegraphics[width=3.0cm]{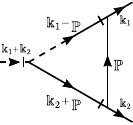}} 
		-2 \raisebox{-1.3cm}{\includegraphics[width=3.0cm]{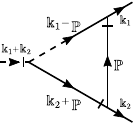}}\\
		%%%%%%%%%%%%%%%%%%%%%%%%%%%%%%%%%%%%%%%%%%%%%%%%%%%%%%%%%%%%%%%%
		& 
		-2\raisebox{-1.3cm}{\includegraphics[width=3.0cm]{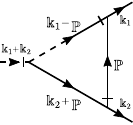}}
		-\raisebox{-1.3cm}{\includegraphics[width=3.0cm]{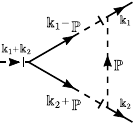}}
		-2\raisebox{-1.3cm}{\includegraphics[width=3.0cm]{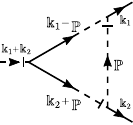}} 
		-2\raisebox{-1.3cm}{\includegraphics[width=3.0cm]{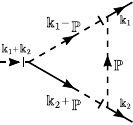}} \\
		%%%%%%%%%%%%%%%%%%%%%%%%%%%%%%%%%%%%%%%%%%%%%%%%%%%%%%%%%%%%%%%%
		&
		-2\raisebox{-1.3cm}{\includegraphics[width=3.0cm]{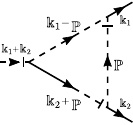}} 
		-2\raisebox{-1.3cm}{\includegraphics[width=3.0cm]{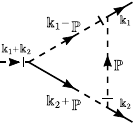}} 
		-2\raisebox{-1.3cm}{\includegraphics[width=3.0cm]{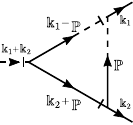}}
		-2\raisebox{-1.3cm}{\includegraphics[width=3.0cm]{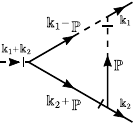}} \\
		%%%%%%%%%%%%%%%%%%%%%%%%%%%%%%%%%%%%%%%%%%%%%%%%%%%%%%%%%%%%%%%%
		&
		-2\raisebox{-1.3cm}{\includegraphics[width=3.0cm]{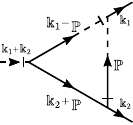}} 
		-2\raisebox{-1.3cm}{\includegraphics[width=3.0cm]{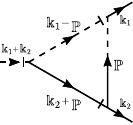}}
		-2\raisebox{-1.3cm}{\includegraphics[width=3.0cm]{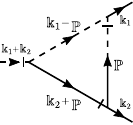}}
		-2\raisebox{-1.3cm}{\includegraphics[width=3.0cm]{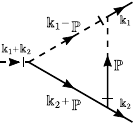}} \\
		%%%%%%%%%%%%%%%%%%%%%%%%%%%%%%%%%%%%%%%%%%%%%%%%%%%%%%%%%%%%%%%%
		& 
		-2\raisebox{-1.3cm}{\includegraphics[width=3.0cm]{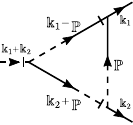}}
		-2\raisebox{-1.3cm}{\includegraphics[width=3.0cm]{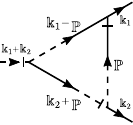}}
		-2\raisebox{-1.3cm}{\includegraphics[width=3.0cm]{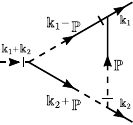}}
		-\raisebox{-1.3cm}{\includegraphics[width=3.0cm]{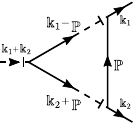}} \\
		%%%%%%%%%%%%%%%%%%%%%%%%%%%%%%%%%%%%%%%%%%%%%%%%%%%%%%%%%%%%%%%%
		& 
		-2\raisebox{-1.3cm}{\includegraphics[width=3.0cm]{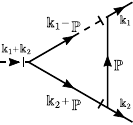}}
		-2\raisebox{-1.3cm}{\includegraphics[width=3.0cm]{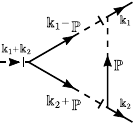}}
		-2\raisebox{-1.3cm}{\includegraphics[width=3.0cm]{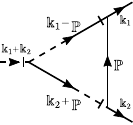}}
		-2\raisebox{-1.3cm}{\includegraphics[width=3.0cm]{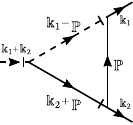}} \\
		%%%%%%%%%%%%%%%%%%%%%%%%%%%%%%%%%%%%%%%%%%%%%%%%%%%%%%%%%%%%%%%%
		&
		-2\raisebox{-1.3cm}{\includegraphics[width=3.0cm]{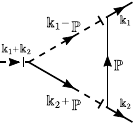}}
		-2\raisebox{-1.3cm}{\includegraphics[width=3.0cm]{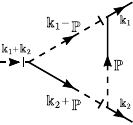}}
		-2\raisebox{-1.3cm}{\includegraphics[width=3.0cm]{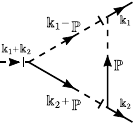}}
		-2\raisebox{-1.3cm}{\includegraphics[width=3.0cm]{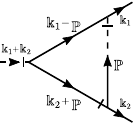}} 
		%%%%%%%%%%%%%%%%%%%%%%%%%%%%%%%%%%%%%%%%%%%%%%%%%%%%%%%%%%%%%%%%
	\end{aligned}
\end{equation}	

\begin{equation}\label{eq:DetH-GPtHH-oneloop2}
	\begin{aligned}
		& 
		%%%%%%%%%%%%%%%%%%%%%%%%%%%%%%%%%%%%%%%%%%%%%%%%%%%%%%%%%%%%%%%%
		-2\raisebox{-1.3cm}{\includegraphics[width=3.0cm]{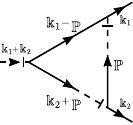}}
		-2\raisebox{-1.3cm}{\includegraphics[width=3.0cm]{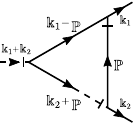}}
		-2\raisebox{-1.3cm}{\includegraphics[width=3.0cm]{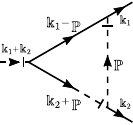}}
		-2\raisebox{-1.3cm}{\includegraphics[width=3.0cm]{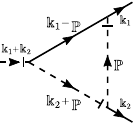}} \\
		%%%%%%%%%%%%%%%%%%%%%%%%%%%%%%%%%%%%%%%%%%%%%%%%%%%%%%%%%%%%%%%%
		& 
		-2\raisebox{-1.3cm}{\includegraphics[width=3.0cm]{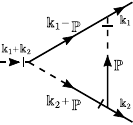}}
		-2\raisebox{-1.3cm}{\includegraphics[width=3.0cm]{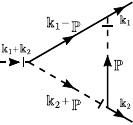}}
		-2\raisebox{-1.3cm}{\includegraphics[width=3.0cm]{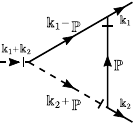}}
		-2\raisebox{-1.3cm}{\includegraphics[width=3.0cm]{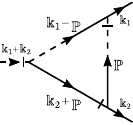}} \\
		%%%%%%%%%%%%%%%%%%%%%%%%%%%%%%%%%%%%%%%%%%%%%%%%%%%%%%%%%%%%%%%%
		& 
		-2\raisebox{-1.3cm}{\includegraphics[width=3.0cm]{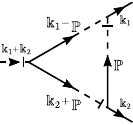}}
		-2\raisebox{-1.3cm}{\includegraphics[width=3.0cm]{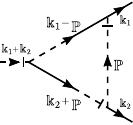}}
		-2\raisebox{-1.3cm}{\includegraphics[width=3.0cm]{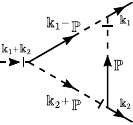}}
		-2\raisebox{-1.3cm}{\includegraphics[width=3.0cm]{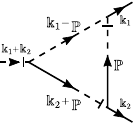}} \\
		%%%%%%%%%%%%%%%%%%%%%%%%%%%%%%%%%%%%%%%%%%%%%%%%%%%%%%%%%%%%%%%%
		& 
		-2\raisebox{-1.3cm}{\includegraphics[width=3.0cm]{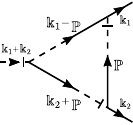}}
		-2\raisebox{-1.3cm}{\includegraphics[width=3.0cm]{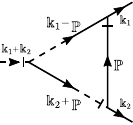}} = \mathcal{O}(k^{4}).
		%%%%%%%%%%%%%%%%%%%%%%%%%%%%%%%%%%%%%%%%%%%%%%%%%%%%%%%%%%%%%%%%
	\end{aligned}
\end{equation}
Surprisingly, the net divergent contribution to $ \Gamma^{\tilde{\phi}hh} $ is zero. As no obvious symmetry forbids the renormalization of this three-point vertex function, we expect this cancellation to be only an artefact of the one-loop approximation. Similar situation occurs in the CKPZ equation as well \cite{Sun89,Caballero18}, where Janssen showed that two-loop contributions, although small, do exists \cite{Janssen97}.

The above UV divergent terms are eliminated by a redefinition of the fields and parameters of the model. As mentioned in the main text, the perturbative corrections are calculated with bare parameters (we only suppress the subscript $ 0 $ for the simplicity). The renormalized quantities are defined similarly as in \eqref{eq:dimensionless-coupling-KPZ}
\begin{align}
	%%%%%%%%%%%%%%%%%%%%%%%%%%%%%%%%%%%%%%%%%%%%%%%%%%%%%%%%%%%%%%%%
	\psi_0 = Z_\psi \psi, \quad \lambda_0 =\frac{\mu^{\frac{4-d}{2}}}{\sqrt{S_d}} Z_\lambda \lambda ,\quad \kappa_0 = \frac{\mu^{\frac{4-d}{2}}}{\sqrt{S_d}} Z_\kappa \kappa,
	%%%%%%%%%%%%%%%%%%%%%%%%%%%%%%%%%%%%%%%%%%%%%%%%%%%%%%%%%%%%%%%%
\end{align}
The renormalization constants are obtained using the minimal subtraction scheme, and the corresponding beta functions have the following form
\begin{align}
	%%%%%%%%%%%%%%%%%%%%%%%%%%%%%%%%%%%%%%%%%%%%%%%%%%%%%%%%%%%%%%%%
	\beta_{\lambda} &= - \lambda \left( \frac{\varepsilon}{2} + \frac{3 \lambda \kappa  (q+1)^2 (3 q-2) - \lambda^{2} \left(c (q+2)-3 q^3+10 q+8\right)}{16 (c+1)^2 (q+1)^3} \right), \\
	%%%%%%%%%%%%%%%%%%%%%%%%%%%%%%%%%%%%%%%%%%%%%%%%%%%%%%%%%%%%%%%%
	\beta_{\kappa} &= - \kappa \left( \frac{\varepsilon}{2} + \frac{3 \lambda \kappa  (q+1)^2 (3 q-4) - \lambda^{2} \left(c (3 q+4)-3 q^3+6 q^2+24 q+16\right) }{16 (c+1)^2 (q+1)^3} \right), \\
	%%%%%%%%%%%%%%%%%%%%%%%%%%%%%%%%%%%%%%%%%%%%%%%%%%%%%%%%%%%%%%%%
	\beta_{q} &= -\frac{q \left(\lambda^{2}  \left(c+q^2+2 q+2\right)+3 \lambda \kappa  (q+1)^2\right)}{8 (c+1)^2 (q+1)^2}, \\
	%%%%%%%%%%%%%%%%%%%%%%%%%%%%%%%%%%%%%%%%%%%%%%%%%%%%%%%%%%%%%%%%
	\beta_{c} &= - \frac{A_{c} \lambda^{2} + B_{c} \lambda \kappa + C_{c} \kappa^{2}}{48(1+c)^{2}(1+q)^{3}},
	%%%%%%%%%%%%%%%%%%%%%%%%%%%%%%%%%%%%%%%%%%%%%%%%%%%%%%%%%%%%%%%%
\end{align}
where
\begin{align}
	%%%%%%%%%%%%%%%%%%%%%%%%%%%%%%%%%%%%%%%%%%%%%%%%%%%%%%%%%%%%%%%%
	A_{c} &= -c^2 (34 q+6)+c \left(6 q^3-15 q^2-106 q-29\right)+q \left(3 q^2+2 q-29\right), \\
	%%%%%%%%%%%%%%%%%%%%%%%%%%%%%%%%%%%%%%%%%%%%%%%%%%%%%%%%%%%%%%%%
	B_{c} &= 2 c (q+1) \left(9 q^2-5 q-9\right) + (q+1) \left(-11 q^2-16 q+5\right), \\
	%%%%%%%%%%%%%%%%%%%%%%%%%%%%%%%%%%%%%%%%%%%%%%%%%%%%%%%%%%%%%%%% 
	C_{c} &= -26 q (q+1)^2.
	%%%%%%%%%%%%%%%%%%%%%%%%%%%%%%%%%%%%%%%%%%%%%%%%%%%%%%%%%%%%%%%%
\end{align}
The anomalous dimensions for fields and $ \nu $ are
\begin{align}
	%%%%%%%%%%%%%%%%%%%%%%%%%%%%%%%%%%%%%%%%%%%%%%%%%%%%%%%%%%%%%%%%
	\gamma_{h} &= - \gamma_{\tilde{h}} = \frac{\lambda^{2} \left(c (q+2)-q^3+4 q^2+12 q+8\right)+ 3 \kappa \lambda (q-2) (q+1)^2 }{16 (c+1)^2 (q+1)^3}, \\
	%%%%%%%%%%%%%%%%%%%%%%%%%%%%%%%%%%%%%%%%%%%%%%%%%%%%%%%%%%%%%%%%
	\gamma_{\phi} &= - \gamma_{\tilde{\phi}} = -\frac{ \lambda^{2} q \left(c+q^2+2 q+2\right) + 3 \kappa \lambda q (q+1)^2 }{16 (c+1)^2 (q+1)^3}, \\
	%%%%%%%%%%%%%%%%%%%%%%%%%%%%%%%%%%%%%%%%%%%%%%%%%%%%%%%%%%%%%%%%
	\gamma_{\nu} &= -\frac{\lambda  q (3 \kappa +\lambda )}{8 (c+1)^2 (q+1)}.
	%%%%%%%%%%%%%%%%%%%%%%%%%%%%%%%%%%%%%%%%%%%%%%%%%%%%%%%%%%%%%%%%
\end{align}

\end{widetext}

\bibliographystyle{apsrev4-1}
\bibliography{ActiveMembrane,FieldTheory}

\end{document}